\documentclass{tlp}

\usepackage{graphicx}
\usepackage{color}

\usepackage[utf8]{inputenc}
\usepackage{stmaryrd}
\usepackage{fancybox}

\usepackage{booktabs}

\usepackage[flushleft]{threeparttable}

\usepackage{algorithm}
\usepackage[noend]{algpseudocode}

\usepackage{amssymb}

\newtheorem{example}{Example}[section]%
\usepackage[T1]{fontenc}

\newcommand{\figskip}{\vspace*{-2mm}}
\newcommand{\underfig}{\vspace*{-4mm}}

\usepackage{xspace}

\newcommand{\ciao}{\texttt{Ciao}\xspace}
\newcommand{\ciaopp}{\texttt{CiaoPP}\xspace}

\usepackage{xcolor}
\usepackage{listings}
\usepackage{textcomp}

\definecolor{ciaoframe}     {rgb}{  0,    0,  0.3}
\definecolor{ciaostring}    {rgb}{0.6, 0.46, 0.33}
\definecolor{ciaooperators} {rgb}{0.1, 0.15,  0.6}
\definecolor{ciaokeywords}  {rgb}{0.1, 0.15,  0.6}
\definecolor{ciaoassertions}{rgb}{0.1, 0.15,  0.6}
\definecolor{ciaotrust}     {RGB}{200, 130,     0}
\definecolor{ciaocheck}     {rgb}{0.1, 0.2,   0.8}
\definecolor{ciaochecked}   {rgb}{0.2, 0.34,  0.1}
\definecolor{ciaotrue}      {rgb}{0.2, 0.34,  0.1}
\definecolor{ciaofalse}     {rgb}{0.6,  0.0, 0.09}
\definecolor{ciaoprops}     {rgb}{0.1,  0.2,  0.8}
\definecolor{ciaocomment}   {rgb}{0.5,  0.5,  0.5}

\newcommand{\prettylstciao}[0]{
\lstset{language=Prolog,
  xleftmargin=3mm,
  frameround=fttt,
  frame=ltrb,
  rulecolor=\color{ciaoframe},
  numbers=left,numberstyle=\tiny,stepnumber=1,numbersep=8pt,
  tabsize=4,
  breaklines=true,breakatwhitespace=true,
  basicstyle=\scriptsize\ttfamily, %
  showlines=true,
  showspaces=false,
  showtabs=false,
  escapechar=@,
  escapeinside={~~},
  commentstyle=\color{ciaocomment},
  stringstyle=\color{ciaostring},
  showstringspaces=false,
  deletekeywords={true}, %
  keywordstyle={\color{ciaooperators}\bfseries}, %
  classoffset=1, %
        otherkeywords={>,<,>=,=<,.,;,-,!,=,*,\&,+,:-,[,],|,->,:,:=,\#},
        keywordstyle={\color{ciaokeywords}\bfseries},
  classoffset=2,
       morekeywords={module,use_module,dynamic,export,import,impl_defined,trait,impl},
       keywordstyle={\color{ciaokeywords}\bfseries},
       morekeywords={pred,prop,calls,success,comp,compat,inst,modedef},
       keywordstyle={\color{ciaoassertions}\bfseries},
  classoffset=4,
       morekeywords={trust,trust_default,entry},
       keywordstyle={\color{ciaotrust}\bfseries},
  classoffset=5,
       morekeywords={check},
       keywordstyle={\color{ciaocheck}\bfseries},
  classoffset=6,
       morekeywords={checked},
       keywordstyle={\color{ciaochecked}\bfseries},
  classoffset=7,
       morekeywords={true},
       keywordstyle={\color{ciaotrue}\bfseries},
  classoffset=8,
       morekeywords={false},
       keywordstyle={\color{ciaofalse}\bfseries},
  classoffset=9,
       morekeywords={even,nat,int,flt,atm,term,num,var,list,ground,mshare,
                    rsize,cardinality,not_fails,fails,exp,cost,costb,steps_ub,steps_lb,
                    size_ub,size_lb,covered,mut_exclusive,head_cost,literal_cost,
                    is_det,det,nondet,semidet,multi,length,terminates,steps_o,resource,socket,seff,string,
                    not_further_inst,nonground,unknown,unreachable,
                    sorted_int_list, sorted, no_choicepoints, shfr, any,
                    leq, builtin
       },
       keywordstyle={\color{ciaoprops}\bfseries},
  classoffset=0, %
}}

\usepackage{lineno}

\renewcommand{\_}{{\fontfamily{ptm}\selectfont\textunderscore}}

\newcommand\tbox[1]{\tikz[overlay]\node[inner sep=2pt, draw=black, thick, anchor=text, rectangle] {#1};\phantom{#1}}
\newcommand\dbox[1]{\tikz[overlay]\node[inner sep=2pt, draw=black, thick, anchor=text, rectangle, dashed] {#1};\phantom{#1}}
\usepackage[normalem]{ulem}

\usepackage[pagebackref=true]{hyperref}

\usepackage{url}

\newcommand{\kbd}[1]{\mbox{\tt #1}}

\renewcommand{\emph}[1]{\textit{#1}}

\usepackage{tikz}
\usetikzlibrary{positioning,shapes,arrows,fit,shadows,calc,matrix}

\usepackage[disable, textsize=scriptsize, colorinlistoftodos, textwidth=1.3cm]{todonotes}
\newcommand{\selfnote}[1] {\todo[backgroundcolor=yellow!20, tickmarkheight=0.1cm]{IC: #1}}
\newcommand{\dfnote}[1] {\todo[backgroundcolor=red!20, tickmarkheight=0.1cm]{DF: #1}}
\newcommand{\mhnote}[1] {\todo[backgroundcolor=blue!20, tickmarkheight=0.1cm]{MH: #1}}

\newcommand{\jfnote}[1] {\todo[backgroundcolor=orange!20, tickmarkheight=0.1cm]{JF: #1}}

\makeatletter 
\@mparswitchfalse%
\makeatother
\normalmarginpar

\begin{document}

\lefttitle{Cambridge Author}

\jnlPage{1}{8}
\jnlDoiYr{2021}
\doival{10.1017/xxxxx}

\title[Checkification: Testing Your (Static Analysis) Truths]
  {Checkification: A Practical Approach for\\
    Testing Static Analysis Truths%
    \thanks{Partially funded by MICINN
    projects PID2019-108528RB-C21 \emph{ProCode}, TED2021-132464B-I00
    \emph{PRODIGY}, and FJC2021-047102-I, and by the Tezos
    foundation.\\ [0.5mm]
    We would also like to thank the anonymous reviewers for their very useful
    feedback.\\ [0.5mm]
    This paper is an extended version of our previous work published in LOPSTR'20.
    }}

\begin{authgrp}
\author{
  DANIELA FERREIRO$^{1,2}$
  IGNACIO CASSO$^{1,2}$ 
  JOSE F. MORALES$^{1,2}$ \\[0.2cm]
  PEDRO LÓPEZ-GARCÍA$^{3,2}$ 
  MANUEL V. HERMENEGILDO$^{1,2}$
  \ \\
  \ \\
  \affiliation{$^1$Universidad Polit\'{e}cnica de Madrid (UPM)}
  \affiliation{\hspace*{-5mm}\email{d.ferreiro@alumnos.upm.es,\{josefrancisco.morales,manuel.hermenegildo\}@upm.es}}
  \vspace{1mm}
  \affiliation{$^2$IMDEA Software Institute}
  \affiliation{\hspace*{-4mm}\email{\{daniela.ferreiro,ignacio.decasso,josef.morales,pedro.lopez,manuel.hermenegildo\}@imdea.org}\hspace*{-4mm}}
  \vspace{1mm}
  \affiliation{$^3$Spanish Council for Scientific Research (CSIC) Madrid, Spain}
}
\end{authgrp}

\history{\sub{17 January 2025;} \rev{22 April 2025;} \acc{30 April 2025}}

\maketitle

\begin{abstract}

Static analysis is an essential component of many modern software
development tools.  
Unfortunately, the ever-increasing complexity of static analyzers
makes their coding error-prone.
Even analysis tools based on rigorous mathematical techniques, such as
abstract interpretation, are not immune to bugs.
Ensuring the correctness and reliability of software analyzers
is critical if they are to be
inserted in production compilers and development environments.
While compiler validation has seen notable success, formal
validation of static analysis tools remains relatively unexplored.
In this paper we present \emph{checkification},
a simple, automatic method for testing
static analyzers.
Broadly, it consists in checking, over a suite of benchmarks, that
the properties inferred statically are satisfied dynamically.
The main advantage of our approach lies in its simplicity, which stems
directly from framing it within the \ciao assertion-based validation
framework, and its blended static/dynamic assertion checking approach.
We demonstrate that in this setting, the analysis 
can be tested %
with little effort by combining the following
components already present in the framework:
1) the \textit{static analyzer}, which outputs its results as the
original program source with assertions interspersed;
2) the assertion \textit{run-time checking} mechanism, which
instruments a program to ensure that no assertion is violated at run time;
3) the \textit{random test case generator}, which generates random
test cases satisfying the properties present in assertion
preconditions; and 
4) the \textit{unit-test framework}, which executes those
test cases.
We have applied our approach to the \ciaopp static analyzer, resulting in
the identification of many bugs with reasonable overhead. Most of these
bugs have been either fixed or confirmed, helping us detect a range of
errors not only related to analysis soundness but also within other
aspects of the framework. 
\vspace*{-2mm}
\end{abstract}

\vspace*{-2mm}
\section{Introduction}
\label{introduction}

Static analysis tools play an important role in 
different stages of the software development
cycle, such as code verification and optimization.
However, building modern analyzers for programming languages presents significant
challenges since these systems are typically large and complex, making them prone to bugs. 
This is a limitation to their applicability in real-life production
compilers and development environments, where they are
used in critical tasks %
that need reassurance about the soundness of the analysis results.

However, the validation of static analyzers is a challenging problem,
which is %
not %
well covered in the literature or by existing tools.
This is probably due to the fact that direct application of formal methods is
not always straightforward with code that is so complex and large,
even without considering 
the problem of having precise specifications to check against
---a clear instance of the classic problem of who checks the checker.
In current practice, extensive testing is the most extended and
realistic validation technique, but it poses some significant challenges
too. Testing separate components of the analyzer misses integration
testing, and designing proper oracles for testing the complete tool is
challenging.

In this paper we propose \emph{checkification}, a simple, automatic,
technique for testing %
static analyzers. %
We believe the approach is general in nature, and can be applied
effectively to a wide class of static analyzers, provided some kind of
run-time checking is feasible for the properties inferred.
Herein we develop the proposal for concreteness in the context of the
\ciao~\citep{hermenegildo11:ciao-design-tplp} logic programming-based,
multi-paradigm language.
The \ciao~programming environment includes \ciaopp, a large and
complex abstract interpretation-based static analysis tool which faces
the specific challenges that we are addressing.
Recently, there has been some interesting
work~\citep{Top_Down_Solver-AFP} aimed at verifying the partial
correctness of the PLAI analysis \emph{algorithm} (also referred to as
``the top-down solver'') that lies at the heart of \ciaopp using the
Isabelle prover~\citep{Paulson90}, but verification of the actual
implementation remains a challenge.
Like other ``classic'' analyzers, the \ciaopp formal framework has evolved for a long
time, incorporating a large number of abstract domains, features, and
techniques, adding up to over half a million lines of code.
These components have in turn reached over the years different levels
of maturity. While the essential parts, such as the fixpoint
algorithms and the classic abstract domains, have been used
routinely for a long time now and it is unusual to find bugs, other
parts are less developed and yet others are prototypes or even proofs of
concept (see Table~\ref{domains} for an overview of some of the
abstract domains that are bundled with the system and their maturity status).
We show in Section~\ref{examples} how %
our proposed method
reveals bugs, not only in the less-developed parts of the system
but also in corner cases of the more mature components,
such as the handling of \textit{built-ins}, run-time checking instrumentation, etc.
A feature
of \ciao that will
be instrumental to our approach is the use of a unified assertion
language %
across
the components of the \ciao
framework~\citep{prog-glob-an,ciaopp-sas03}, %
which together implement a unique %
blend of static and dynamic assertion
checking. These components (and their algorithms) include:
\begin{enumerate}
\item The \emph{static
  analyzer}~\citep{ai-jlp,incanal-toplas,incanal-assrts-openpreds-lopstr19-post},
(top-down analysis framework) %
which expresses the inferred information as assertions
interspersed within the original program.
\item The assertion \emph{run-time checking framework}~\citep{cached-rtchecks-iclp2015,optchk-ppdp2016},
which instruments the code to ensure that any assertions remaining
after static verification are not violated at run time.
\item The \emph{(random) test case} generation
framework~\citep{testgen-lopstr19-post}, which generates random test
cases satisfying the properties present in assertion preconditions.
\item The \emph{unit-test framework}~\citep{testchecks-iclp09}, which
executes those test cases.
\end{enumerate}

In this paper, we propose an algorithm that combines these four basic
components in a novel way that allows testing the static analyzer
almost for free. %
Intuitively,
it consists in checking, over a suite of benchmarks, that
the properties inferred statically are satisfied dynamically.
The overall testing process, for each benchmark, can be summarized as
follows:
first, the code is analyzed, obtaining the analysis results expressed
as assertions interspersed within the original code.
Then, %
the \emph{status} of these assertions is \emph{switched} into \textit{run-time checks}, that will
 ensure that violations of those assertions are reported at execution time.
Finally, random test cases are generated and executed to
exercise those \textit{run-time checks}.
Given that these assertions (the analyzer output) must cover all
possible concrete executions (and assuming the correctness of our
checking algorithm implementation), if any assertion violation is
reported, assuming that the run-time checks are correct, it means that
the assertion was
incorrectly inferred by the analyzer, thus revealing an error in the
analyzer itself. The error can of course sometimes also be in the
run-time checks,
but typically run-time checking is simpler than inference. 
This process %
is automatable, and, if it is repeated for an
extensive and varied enough suite of benchmarks, it can be used to
effectively validate (even if not fully verify) the analyzer or to
discover new bugs.
Furthermore, the implementation, when framed within a tool environment
that follows \ciao assertion model,
is comparatively simple, at least conceptually.

The idea of checking at run time
the properties or assertions inferred
by the analysis for different program points is not new.
For example, \cite{10.1145/2491411.2491439} successfully applied this
technique for checking a range of different aliasing analyses.
However, these approaches
require the development of tailored instrumentation or monitoring, and 
significant effort in their design and implementation.
We argue that the testing approach is made more applicable, general,
and scalable by the use of a unified assertion-based framework for static
analysis and dynamic debugging, as the \ciao assertions model.
As mentioned before, by developing %
the approach within such a framework,
it can be implemented with many of the already existing
algorithms and components
in the system, in a very simple way. As a result,
our initial prototype was
quite simple, even if, inevitably, the current working version has of
course grown quite a bit in order to add functionality, make it easier
to use, include specific instrumentation, collect performance data,
etc.
Moreover, the components and algorithms of the \ciao system used by
our implementation have been extensively validated,
providing greater confidence in the correctness of our
approach. Consequently, when our method
flags a runtime checking error, we can be more certain that it
identifies an actual error in the analyzer. If no error exists in the
analyzer, such runtime checking errors help us locate and fix bugs in
our implementation. These fixes will likely address bugs in other
\ciao system components used by our implementation. Importantly, these
components can also serve other purposes in the software development
process. In conclusion, any runtime checking error flagged by our
approach contributes to improving the entire \ciao system.

We also argue that our approach is particularly useful in a mixed
production and research setting like that of \ciaopp, in which there
is a mature and domain-parametric abstract interpretation framework
used routinely, but new, experimental abstract domains and overall
improvements are in constant development.
Those domains can easily be tested relying only on the existing
abstract-interpretation framework, run-time checking framework, and
unified assertion language, provided only that the assertion language
is extended to include the properties %
that are 
inferred by the domains.

The rest of the paper is structured as follows:
Section~\ref{sec:preliminaries} provides the background needed for
describing the main ideas and contributions of the paper. In
particular, we describe the basic components used in our approach.
Section~\ref{algorithm} then presents and discusses our proposed
``checkification'' algorithm for testing static analyzers, with an
initial illustrative example, the basic reasoning behind the approach
(Section~\ref{sec:basicreasoning}), the operation of the algorithm
(Section~\ref{sec:operation}), and discussions of some of additional
issues involved (Sections~\ref{sec:properties}
and~\ref{sec:multivariance}).
In Section~\ref{evaluation} we present our experimental
evaluation and results. We explain the evaluation setup
(Section~\ref{experiments}) including experiments, analyzer
configuration, abstract domains and properties studied, and programs
analyzed. We then present and discuss the results of this evaluation
(Section~\ref{sec:results}), in terms of the errors found and cost of
the technique.  Section~\ref{examples} then presents further
discussion with examples of the classes of errors detected which also
serves to go over some of the practical uses
of the approach.
We conclude by discussing additional related work in
Section~\ref{sec:related-work}, and presenting some conclusions and
perspectives in Section~\ref{conclusions}.

\vspace*{-6mm} 
\section{Basic Components}  
\label{sec:preliminaries}

\paragraph{\textbf{Assertion Language.}}

Assertions are syntactic objects which allow expressing
properties of programs that should hold at certain points of program execution.
Assertions are used everywhere in \ciao, from documentation and foreign
interface definitions to static analysis and dynamic debugging.
Two types of \ciao assertions that are relevant herein are 
\textit{predicate} assertions (\texttt{pred} for short) and
\textit{program-point} assertions:%
\footnote{We will also use an additional form, \texttt{entry} assertions, that will be introduced later.}
The first ones are declarations that provide partial specifications of
a predicate.
They have the following syntax:\\ 
\centerline{$ \kbd{:- [}\textit{Status} \kbd{] pred } \textit{Head} \kbd{ [: } \textit{Calls} \kbd{] [=> } \textit{Success} \kbd{] [+ } \textit{Comp} \kbd{].} $}

\smallskip
\noindent
and express that a) calls to predicate \textit{Head} that satisfy
precondition \textit{Calls} are admissible and b) that, for such calls, 
the predicate must satisfy post-condition \textit{Success}
if it succeeds, and global computational properties \textit{Comp}.
If there are several \texttt{pred} assertions, the set of
\textit{Calls} fields define the admissible calls to the predicate.
\textit{Program-point} assertions are reserved literals that appear in
the body of clauses and describe properties that hold in the
run-time constraint store every time execution reaches that point in
the clause at run time.
Their syntax is
\textit{Status}\kbd{(}\textit{State}\kbd{)}.
Both of these kinds of assertions can have different values in the
\textit{Status} field depending on their origin and intended use.
Assertion statuses relevant herein include:
\begin{itemize}
\item \texttt{true}, which is the status of the assertions that are
  output from the analysis (and thus
  must be
  safe approximations of the concrete semantics);
\item \texttt{check}, which indicates that the validity of the
  assertion is unknown and it must be checked, statically or
  dynamically, and is the \emph{default} value of \textit{Status} when
  not indicated; and,
\item \texttt{trust}, which indicates that the
  analyzer should assume this assertion to be correct, even if it
  cannot be automatically inferred.
\end{itemize}

\vspace*{-3mm} 
\begin{example}[Some assertions]
The following code fragment provides examples of both types of
assertions, predicate and program-point; all these assertions have
status \texttt{check}:

\prettylstciao
\begin{lstlisting}
  :- check pred append(X,Y,Z) : (list(X),list(Y),var(X)) => list(Z) + det.
  :- check pred append(X,Y,Z) : (var(X),var(Y),list(Z)) => (list(X),list(Y)) + multi.
    
  append([],X,X).
  append([X|Xs],Ys,[X|Zs]) :-
    append(Xs,Ys,Zs),
    check(list(Xs),list(Ys),list(Zs)).
\end{lstlisting}
The first two \texttt{pred} assertions define two different ways in
which \texttt{append/3} is expected to be called. The first one
states that \texttt{append/3} may be called with the two
first arguments instantiated to lists and the third a variable, and
that, if such a call succeeds, then the third argument should be
bound to a list. This first assertion also states that when called
like this, the predicate should have only one solution and should not
fail (\texttt{det}, a global computational property). The second
\texttt{pred} assertion states that \texttt{append/3} may also be
called with the third argument instantiated to a list and the first
two variables, and that, if such a call succeeds, then the first and
second arguments should be bound to lists, and that in this case the
predicate should produce one or more 
solutions (\texttt{multi}, also a global computational property), but, again, not fail.
There is also a program-point assertion in the second
clause of \texttt{append/3} that states that if execution reaches that
point in that clause, all of \texttt{Xs}, \texttt{Ys}, and \texttt{Zs}
should be bound to lists.
For all these assertions the \texttt{check} status indicates that
these are desired properties that need to be checked, statically or
dynamically, but have not been proven true or false yet.
\end{example}

Assertion fields \textit{Calls}, \textit{Success}, \textit{Comp} and
\textit{State} are conjunctions of %
\emph{properties}.
Such properties are predicates, typically written in the source
language (user-defined or in libraries), and thus runnable, so that
they can be used as run-time checks. For our purposes herein, we will
consider typically properties that are \emph{native} to \ciaopp, i.e.,
that can be abstracted and inferred by some domain in \ciaopp.
This includes a wide range of properties, from types, modes and
variable sharing, to determinacy, (non)failure, and resource
consumption.
We refer the reader to~\cite{assert-lang-disciplbook,
  ciaopp-sas03-journal-scp, hermenegildo11:ciao-design-tplp} and their
references for a full description of the \ciao~assertion language.

In the \ciao assertion syntax, properties can also be \textit{in-lined} in
the predicate arguments, also referred to as using \textit{modes}.  Such
modes are \emph{property macros} that serve to specify in a compact
way several properties referring to a predicate argument.  A specific
syntax, resembling that of \textit{predicate} assertions is used to
define modes.

\begin{example}[Modes]
  \label{exa:modes}
For example, if the following modes are
defined:\footnote{Note that ``\texttt{-}'' is often also defined
  simply as: \texttt{:- modedef -(A,P) => P(A).} As mentioned before, in 
  Ciao modes are user-definable.}
\prettylstciao
\begin{lstlisting}
:- modedef  +(A,P) : P(A).           % A has property P in calls
:- modedef  -(A,P) : var(A) => P(A). % A is var on calls and has property P on success
\end{lstlisting}

\noindent
then the assertions in the previous program can be expressed equivalently
as follows: 
\prettylstciao
\begin{lstlisting}
  :- check pred append(+list,+list,-list) + det.
  :- check pred append(-list,-list,+list) + multi.
\end{lstlisting}
\vspace*{-3mm}
\end{example}

\begin{figure}[t]
  \centering
  \resizebox{\textwidth}{!}{
    \hspace*{-20mm}
\pgfdeclarelayer{background}
\pgfdeclarelayer{foreground}
\pgfsetlayers{background,main,foreground}

\tikzstyle{source}=[draw, draw=cyan!80!black!100, fill=cyan!20, text width=6em, font=\sffamily,
    thick,
    minimum height=2.5em,drop shadow]
\tikzstyle{tool}=[draw=green!50!black!100, fill=green!40, text width=5em, font=\sffamily, 
    thick,
    text centered, 
    chamfered rectangle, chamfered rectangle angle=30, chamfered rectangle xsep=2cm]
\tikzstyle{newtool}=[draw=yellow!50!black!100, fill=yellow!40, text width=5em, font=\sffamily, 
    thick,
    text centered, 
    chamfered rectangle, chamfered rectangle angle=30, chamfered rectangle xsep=2cm]
\tikzstyle{midresult}=[draw, fill=white!40, text width=5em, font=\sffamily,
    thick,
    rounded rectangle,
    minimum height=1em,drop shadow]
\tikzstyle{warnresult}=[color=orange!50!black!100, fill=orange!40, text width=6em, font=\sffamily, 
    thick,
    rounded corners=2pt,
    minimum height=1em,drop shadow]
\tikzstyle{errresult}=[color=red!80!black!100, fill=red!20, text width=6em, font=\sffamily, 
    thick,
    rounded corners=2pt,
    minimum height=1em,drop shadow]
\tikzstyle{okresult}=[color=green!50!black!100, fill=green!40, text width=6em, font=\sffamily, 
    thick,
    rounded corners=2pt,
    minimum height=1em,drop shadow]
\tikzstyle{certresult}=[color=blue!50!black!100, fill=blue!40, text width=5em, font=\sffamily, 
    thick,
    rounded corners=2pt,
    minimum height=1em,drop shadow]
\tikzstyle{coderesult}=[color=blue!50!black!100, fill=blue!40, text width=5em, font=\sffamily, 
    thick,
    rounded corners=2pt,
    minimum height=1em,drop shadow]

\scriptsize
\begin{tikzpicture}[>=latex]
  \node (code) [source] {
    \textbf{Code}\\
    (\mbox{user, builtins,} libraries)
  };
  \path (code)+(0,-7em) node (assertions) [source] {
    \textbf{Assertions}\\
    (\mbox{user, builtins,} libraries)\\[1ex]
    :- check\\
    :- test\\
    :- trust\\
    Unit-tests \\
  };

\path (code)+(10em,2em) node (statana) [tool] {Static Analysis (Fixpoint)};
\path (statana)+(0,-12em) node (normalizer) [tool] {Assertion Normalizer \& Library Interface};
\path (statana)+(9.5em,0) node (anainfo) [source] {Analysis Info};
\path (anainfo)+(0em,-12em) node (comparator) [tool] {Static Comparator \& Simplifier};
\path (anainfo)+(9em,0em) node (rtcheck) [tool] {RT-Check};
\path (rtcheck)+(-3em,-2.8em) node (unittest) [tool] {Unit-Test};
\path[color=brown!50!black] (comparator)+(6em,4em) node (texec) [midresult] {:- texec};
\path[color=blue] (comparator)+(9em,2em) node (check) [midresult] {:- check};
\path[color=red] (comparator)+(9em,-0em) node (false) [midresult] {:- false};
\path[color=green!50!black] (comparator)+(9em,-2em) node (checked) [midresult] {:- checked};
\path[color=green!50!black] (comparator)+(-0em,+3.5em) node (true) [midresult] {:- true};
\path (rtcheck)+(8.5em,0) node (rterror) [errresult] {Possible\\run-time error};
\path (false)+(8.5em,1em) node (cterror) [errresult] {Compile-time error};
\path (check)+(8.5em,2em) node (verifwarn) [warnresult] {Verification warning};
\path (checked)+(8.5em,0) node (verified) [okresult] {Verified};

\path (unittest)+(2em,-2.7em) node (generator) [newtool] {Test Case \\ Generator};

\path [draw, thick, ->] (code.east) -- node [] {} (statana.west) ;
\path [draw, thick, ->] (assertions) -- node [] {} (normalizer.west) ;
\path [draw, thick, ->] (normalizer) -- node [] {} (statana) ;
\path [draw, thick, ->] (normalizer.east) -- node [] {} (comparator.west) ;
\path [draw, thick, ->] (comparator.east) -- node [] {} (texec.south west) ;
\path [draw, thick, ->] (comparator.east) -- node [] {} (check.west) ;
\path [draw, thick, ->] (statana) -- node [] {} (anainfo) ;
\path [draw, thick, ->] (comparator.east) -- node [] {} (false.west) ;
\path [draw, thick, ->] (comparator.east) -- node [] {} (checked.west) ;
\path [draw, thick, ->] (anainfo) -- node [] {} (true) ;
\path [draw, thick, ->] (true) -- node [] {} (comparator) ;
\path [draw, densely dashed, thick, ->] (true) -- node [] {} (check) ;
\path [draw, densely dashed, thick, ->] (anainfo) -- node [] {} (rtcheck) ;
\draw [thick, ->] (texec.north west) to [bend left] (unittest.south west) ;
\path [draw, thick, ->] (unittest) -- node [] {} (rtcheck) ;
\path [draw, thick, ->] (rtcheck) -- node [] {} (rterror) ;
\draw [thick, ->] (check.north east) to [bend right] (rtcheck) ;
\path [draw, thick, ->] (check) -- node [] {} (verifwarn) ;
\path [draw, thick, ->] (false) -- node [] {} (cterror) ;
\path [draw, thick, ->] (checked) -- node [] {} (verified) ;

\path [draw, thick, ->] (check) to [bend right] (generator) ;
\path [draw, thick, ->] (generator.west) to [bend right] (texec) ;

\path [color=black] (anainfo.north)+(0,1.5em) node (preprocessor) {\sffamily \small Preprocessor};

\path [color=black] (code.north)+(0,1.5em) node (program) {\sffamily \small Program};

\begin{pgfonlayer}{background}
  \path (statana.north west)+(-0.5,1.0) node (g) {};
  \path (checked.south east)+(0.5,-0.5) node (h) {};
  
  \path[fill=yellow!20,rounded corners, draw=black!50, densely dashed] (g) rectangle (h);
\end{pgfonlayer}

\begin{pgfonlayer}{background}
  \path (code.north west)+(-0.1,0.8) node (g) {};
  \path (assertions.south east)+(0.1,-0.1) node (h) {};
  
  \path[source] (g) rectangle (h);
\end{pgfonlayer}

\end{tikzpicture}

}
  \figskip%
  \caption{The \ciao\ assertion framework (\ciaopp's
    verification/testing architecture).} 
  \label{fig:integrated-framework-ciaopp}
\underfig%
\end{figure}

\bigskip
Fig.~\ref{fig:integrated-framework-ciaopp} depicts the overall
architecture of the \ciao\ unified assertion framework. Hexagons
represent tools, and arrows indicate the communication paths among
them. Most of this
communication is performed in terms of assertions.
The input to the process is the user program, optionally
including some assertions. 
Such assertions always include any assertions available in the
libraries for \textit{built-ins}
(the basic operations of the source language),
or for predicates exported by such
libraries that are used by the code being analyzed (left part of
Fig.~\ref{fig:integrated-framework-ciaopp}).

\paragraph{\textbf{Static Program Analysis.}}
\label{ciao-static-analysis}

Abstract interpretation is a formal framework for static analysis that allows
inferring program properties that hold for all possible
program executions. Different abstractions, called abstract domains,
are used in this process for approximating sets of concrete run-time states.
The \ciaopp analyzer is
abstract interpretation-based, and its design consists
of a common abstract-interpretation framework based on fixpoint computation
parameterized by different, \textit{pluggable} abstract domains
(the \textit{Static Analysis}
hexagon in Fig.~\ref{fig:integrated-framework-ciaopp}).
This means that the set of properties that are used in assertions is
extensible with new abstract domains defined as \textit{plug-ins} to support them.
Depending on the selected domain or combination of domains
for analysis, \ciaopp constructs a \textit{program analysis graph},
starting from the program \emph{entry points}. 
In this graph, nodes represent the different ways in which predicates
are called. A predicate can have multiple nodes associated with it
if it is called in different ways (calling contexts).
For each calling context,
properties are inferred that hold if the predicate succeeds (and also
global properties).
These properties will be emitted also as assertions,
which will have status  \texttt{true} 
(represented by the \textit{Analysis Info} box in Fig.~\ref{fig:integrated-framework-ciaopp}).
Optionally, a new source file is generated for the analyzed program, 
which is identical to the original but with \texttt{true}
\textit{program-point} assertions interspersed between every two
consecutive literals of each clause, and with one or more \texttt{true}
\textit{predicate} assertions added for each predicate.
In particular, if there are several different calling contexts for a
given predicate, there will typically be a predicate assertion added
for each of these contexts (this is also referred to
as \emph{multivariance} in the analysis). 
Further details of this process can be found
in~\cite{manuel:tr89,mcctr-fixpt,ai-jlp}.  In any case,
when discussing the identified bugs later in the paper, we will
provide more %
explanations as needed.

\paragraph{\textbf{Run-time Checking.}}

Static analysis is used for compile-time checking of
assertions. However, due to the inherent undecidability of static
analysis, sometimes properties cannot be verified statically.
In those cases, the remaining unproved (parts of) assertions are written into the
output program with \texttt{check} status and then this output program can (optionally) be
instrumented with run-time checks to make it run-time safe.
These dynamic checks will encode the semantics of
the \texttt{check} assertions, ensuring that an error is reported at
run time
if any of these remaining assertions is violated (the dynamic part of
the \ciao assertion model).
Note that almost all current abstract interpretation systems assume in
their semantics that the run-time checks will always be executed.
However, \ciaopp does not make this assumption by default, i.e., it is
configurable as an option, since in some use cases run-time checks may
in fact be disabled by the user for deployment.

Checking at run time program state properties, such as traditional
types and modes, can be performed relatively easily: as mentioned
before, most properties are \textit{runnable}, and the
\texttt{check/1} wrapper will ensure that the check will succeed or
raise an error, without binding any arguments. For example, calling
\texttt{check(list(X))} with \texttt{X = []}, \texttt{X = [a]}, or
\texttt{X = [A,B]},
will succeed, without binding any variables,
while calling \texttt{check(list(X))} with \texttt{X = a}, \texttt{X =
  f(a)}, or \texttt{X = A},
will raise an error.%
\footnote{A discussion of \emph{instantiation} checks and
  \emph{compatibility} checks is appropriate at this point but beyond
  the scope of the paper. Checks are \emph{instantiation} checks
  unless otherwise stated. The reader is referred
  to~\citep{prog-glob-an,ciaopp-sas03} for details.}
In practice a quite rich set of properties is checkable,
including types, modes, variable sharing,
exceptions,
determinacy, (non-)failure, choice-points, and more, blending smoothly
static and dynamic techniques. 
On the other hand, checking at run time other global properties such as
cost and, specially, termination, is obviously less straightforward.
While checking these types of properties could conceptually be
done with our proposed algorithm, in this paper we concentrate on the
other properties mentioned.

\paragraph{\textbf{Unit Tests, Test Case Generation, and Assertion-based Testing.}}

Test inputs can be provided by the user, by means of \texttt{test}
assertions (unit tests). The run-time checking mechanism can test
these assertions but also any
other assertion in any predicate called by the test case, that was not
verified in the static checking. 
The unit-testing framework in principle requires the user to manually write
individual test cases for each assertion to be tested.
However, the~\ciao model also includes mechanisms for generating test
cases automatically from the assertion preconditions, using the
corresponding property predicates as generators. For example, calling
\texttt{list(X)} with \texttt{X} uninstantiated generates lazily,
through backtracking, an infinite set of lists, \texttt{X = [], X =
  [\_], x = [\_,\_],} etc.
Stating a type for the list argument
will then also generate concrete values for the list elements.
This enumeration process can be combined in \ciao with the 
different supported search rules (breadth-first, iterative deepening,
random search, etc.) to produce, e.g., fair enumerations.
This idea has been extended
recently~\citep{testgen-lopstr19-post} to a full random test
case generation framework, %
which automatically generates, using the same technique, \emph{random}
test cases that satisfy assertion preconditions.
We refer to the combination of this test generation mechanism with the
run-time checking of the intervening assertions as
\textit{assertion-based testing}. In other words,
\textit{assertion-based testing} involves generating and running
relevant test cases that exercise the run-time checks of
the assertions in a program to test if those assertions are correct.
This technique (present in the \ciao model since its origins)
yields similar results to \textit{property-based
testing}~\citep{Claessen:2000:QLT:351240.351266-icfp}
but in a more integrated way within the overall assertion model and
within \ciaopp, rather than as a separate technique.
Such automatic generation is currently supported for \emph{native}
properties, \textit{regular types}, and user-defined properties as
long as they are restricted to pure Prolog with arithmetic or mode and
sharing constraints.
In addition, users can also write their own generators and of course
other test generation techniques and tools can be used~\citep{vidal-concolic-testing-tool}.

\section{The Checkification Algorithm}
\label{algorithm}
This section provides a detailed overview of the proposed algorithm
for testing the static analyzer, which incorporates all the 
components mentioned above.

\paragraph{\textbf{Illustrative example.}}
Let us start by sketching the main idea of
our approach with a motivating example.
Assume we have the following simple Prolog program, where we use an 
\texttt{entry}
assertion to define the entry point for the analysis. The
\texttt{entry} assertion indicates that the predicate is %
called with its second argument instantiated to a list and the third a
free variable (we use the mode definitions of Example~\ref{exa:modes}):

\prettylstciao
\begin{lstlisting}
  :- entry prepend(_,+list,-).

  prepend(X,Xs,Ys) :-
      Ys=[X|Rest],
      Rest=Xs.
\end{lstlisting}

\noindent
Assume that we analyze it with a \textit{simple modes} abstract
domain that assigns to each variable in an abstract substitution one of the
following abstract values:

\begin{itemize}
\item %
  \texttt{ground}   (the variable is ground),
\item %
  \texttt{var} (the variable is free),
\item %
  \texttt{nonground} (the variable is not ground),
\item %
  \texttt{nonvar} (the variable is not free),
\item \texttt{ngv}
  (the variable is neither ground nor free), or
\item \texttt{any}
  (nothing can be said about the variable).
\end{itemize}
Assume also that the analysis is incorrect because it does not
consider sharing (aliasing) between variables,
so when updating the
abstract substitution after the \texttt{Rest=Xs} literal,
the abstract value for \texttt{Ys} is not modified at all.%
\footnote{Note that early LP analyzers often had errors of this kind,
  which led to very active development of \emph{variable sharing}
  analysis domains. These constituted some of the very first Abstract
  Interpretation-based pointer aliasing analyses for any programming
  language.}
The result of the analysis will be represented, as explained in the
previous section, as
a new source file with interspersed assertions, as shown
in~Fig.~\ref{fig:modesexample}
(lines~\ref{predbeg}-\ref{predend}, \ref{trueone}, \ref{truetwo}, and
\ref{truethree}).
Note that the correct result, if the analysis considered aliasing,
would be that there is no groundness information for \texttt{Ys} at
the end of the clause (line~\ref{truethree}), since there is none for
\texttt{X} or \texttt{Xs} at the beginning either.
\texttt{Ys} could only be inferred to be \texttt{nonvar}, but instead
is incorrectly inferred to be \texttt{nonground} too (line~\ref{truetwo}).
Normally %
\texttt{any/1} properties (i.e., top, or unknown) would not actually
be included in the analysis output for conciseness, but are included
in~Fig.~\ref{fig:modesexample} for clarity. %

\begin{figure}[t]
\prettylstciao
\begin{lstlisting}[escapechar=@]
  :- entry prepend(_,+list,-).

  :- true pred prepend(X,Xs,Ys)@\label{predbeg}@
      :  (any(X), nonvar(Xs), var(Ys))
      => (any(X), nonvar(Xs), nonground(Ys), nonvar(Ys)).@\label{predend}@

  prepend(X,Xs,Ys) :-
      true(any(X), nonvar(Xs), var(Ys), var(Rest)),@\label{trueone}@
      Ys=[X|Rest],
      true(any(X), nonvar(Xs), nonground(Ys), nonvar(Ys), var(Rest)),@\label{truetwo}@
      Rest=Xs,
      true(any(X), nonvar(Xs), nonground(Ys), nonvar(Ys), nonvar(Rest)).@\label{truethree}@
\end{lstlisting}
  \figskip%
\caption{An incorrect simple mode analysis.}    
\label{fig:modesexample}
\underfig%
\end{figure}
The objective of our approach is to check 
dynamically the validity of these \texttt{true} assertions from the
analyzer, that in this case contain an error.
The insight is that, thanks to the different capabilities of the
\ciao model presented previously,
this can be achieved by (\textbf{1}) \emph{turning the status of the
\texttt{true} assertions produced by the analyzer into
\texttt{check}}, as shown in
Fig.~\ref{fig:instrumented}. \footnote{Again, we include the
\texttt{any/1} property in Fig.~\ref{fig:instrumented} for clarity of
exposition, and for consistency with Fig.~\ref{fig:modesexample}.}
This would normally not make any sense
since these \texttt{true} assertions have been proved by the
analyzer. But that is exactly what we want to check, i.e., whether the
information inferred is incorrect. To do this, (\textbf{2}) we run the
transformed program (Fig.~\ref{fig:instrumented}) again through
\ciaopp (Fig.~\ref{fig:integrated-framework-ciaopp}) but \emph{without
performing any analysis}.
In that case, the \texttt{check} literals
(stemming from the \texttt{true} literals of the previous run) 
will not be simplified in the comparator (since there is no
abstract information to compare against) and instead will be converted
directly to run-time tests. In other words, the \kbd{check(}\textit{Goal}\kbd{)} literals will be
expanded and compiled to code that, every time that this program point is
reached, in every execution, will check dynamically if the property
(or properties) within the \texttt{check} literal (i.e., those in
\textit{Goal}) succeed, and an error message will be emitted if they do not.
\begin{figure}[t]
\prettylstciao
\begin{lstlisting}[escapechar=@]
  :- entry prepend(_,+list,-).

  :- check pred prepend(X,Xs,Ys)
      :  (any(X), nonvar(Xs), var(Ys))
      => (any(X), nonvar(Xs), nonground(Ys), nonvar(Ys)).

  prepend(X,Xs,Ys) :-
      check(any(X), nonvar(Xs), var(Ys), var(Rest)),
      Ys=[X|Rest],
      check(any(X), nonvar(Xs), nonground(Ys), nonvar(Ys), var(Rest)),
      Rest=Xs,
      check(any(X), nonvar(Xs), nonground(Ys), nonvar(Ys), nonvar(Rest)).@\label{line:incorrassert2}@
\end{lstlisting}
  \figskip%
\caption{The instrumented program.}    
\label{fig:instrumented}
\underfig%
\end{figure}
The only missing step to complete the automation of the approach is to
(\textbf{3}) run \texttt{prepend/3} on a set of test cases. These may
in general be already available as test assertions in the program or, 
alternatively, the random test case generator can be used to generate them.
E.g., for \texttt{prepend/3} the test generation framework will ensure
that instances of the goal \texttt{prepend(X,Xs,Ys)} are generated,
where \texttt{Xs} is constrained to be a list, and \texttt{Ys}
remains a free variable.
However, \texttt{X} and the elements of \texttt{Xs} will otherwise be
instantiated to random terms.
In this example, as soon as a test case is generated where both
\texttt{X} and all elements in \texttt{Xs} are ground, the program
will report a run-time error in the \texttt{check} in
line~\ref{line:incorrassert2}, letting us know that the third
program point, and thus the analysis, is incorrect.%
\footnote{In the discussion above we have assumed for simplicity that
  the original program did not already contain \texttt{check}
  assertions. In that case these need to be treated separately and
  there are several options, including simply ignoring them for the
  process or actually turning them into \texttt{trust}s (assertions to
  be taken as granted by the analyzer), so that we
  switch roles and trust the user-provided properties while checking
  the analyzer-inferred ones.
  This very interesting issue of when and whether to use the
  user-provided assertions to be checked during analysis, and its
  relation to run-time checking is discussed in depth
  in~\cite{guided-analysis-post-lopstr18}.}

The same procedure can be followed to debug different analyses with
different benchmarks.
If the execution of any test case reports a run-time error for
one assertion, it will mean that the assertion was not correct and the
analyzer computed an incorrect over-approximation of the semantics of
the program.
Alternatively, if this experiment, which can be automated easily, is
run for an extensive suite of benchmarks without errors, we can gain
more confidence that our analysis implementation is correct, even if
perhaps imprecise (although of course, we cannot have actual 
correctness in general by testing).

\subsection{Basic Reasoning Behind the Approach}
\label{sec:basicreasoning}

We start by establishing more concretely the basic reasoning behind
the approach in terms of abstract interpretation and safe upper and
lower approximations.
The mathematical notation in this subsection is %
meant for providing a more precise explanation, rather than deep
formalization, which is
arguably not really necessary, thanks to the simplicity of the
approach that builds on the different parts of the system that act as
trusted base.

An abstract interpretation-based static analysis computes an
over-approximation $S_P^+$ of the collecting semantics $S_P$ of a
program $P$.
Such collecting semantics can be broadly defined as a control flow
graph for the program decorated at each node with the set of all
possible states that could occur at run time at that program point.
Different approximations of this semantics will have smaller or
larger sets of possible states at each program point.
Let us denote by %
$S'_P \subset_P S''_P$ 
the relation that establishes that an approximation of $S_P$, $S''_P$,
is an over-approximation of another, $S'_P$.
The analysis will be correct if indeed $S_P \subset_P S^+_P$.

Since $S_P$ is undecidable, this relation cannot be checked in
general. However, if we had a good enough under-approximation $S^-_P$
of $S_P$, it can be tested as $S^-_P \subset_P S^+_P$.
If it does not hold and $S^-_P \not\subset_P S^+_P$, then it would
imply that $S_P \not\subset_P S^+_P$, and thus, the results of the
analysis would be incorrect, i.e., the computed $S_P^+$ would not
actually be an over-approximation of $S_P$.

An under-approximation of the collecting semantics of $P$ is easy to
compute: it suffices with running the program with a subset $I^-$ of
the set $I$ of all possible initial states.
We denote the resulting under-approximation $S^{I^-}_P$, and note that
$S_P=S^I_P$, which would be computable if $I$ is finite and $P$
always terminates.
That is the method that we propose for testing the analysis: selecting
a large and varied enough $I^-$, computing $S^{I^-}_P$ and checking
that $S^{I^-}_P \subset_P S^+_P$.

A direct implementation of this idea is challenging.
It would require tailored instrumentation and monitoring to build and
deal with a partially constructed collecting semantic 
under-approximation as a programming structure, which then would need
to be compared to the one the analysis handles.
However, as we have seen the process can be greatly simplified 
by reusing some of the components already in the system,
following these observations:

\begin{itemize}

  \item We can work with one initial state $i$ at a time, following
    this %
    reasoning:\\ $S^{I^-}_P \subset_P S^+_P \iff \forall i
    \in I^-, ~ S^{\{i\}}_P \subset_P S^+_P$.

  \item We can use the random test case generation framework for
    selecting each initial state $i$.
    
  \item Instead of checking $S^{\{i\}}_P \subset_P S^+_P$, we can
    instrument the code with run-time checks to ensure the execution
    from initial state $i$ does not contradict the analysis at any
    point.  That is, to make sure that the state of the program at any
    program point is contained in the over-approximation of the set of
    possible states that
    the analysis inferred and output as \ciao assertions.
\vspace*{-3mm}
\end{itemize}

\subsection{Operation of the Algorithm}
\label{sec:operation}

\begin{algorithm}[t]
\caption{The ``Checkification'' Analysis Testing Algorithm}
\label{alg:anatest}
\begin{algorithmic}[1]
\Procedure{AnaTest}{$P,D$}   \Comment{For program $P$ and domain $\mathcal{D}$}
  \State $result \leftarrow$ \textsc{None}
  \State $P_{an} \leftarrow $ analyze and annotate $P$ with domain $\mathcal{D}$ (incl. program-point assertions).
  \State $P_{check} \leftarrow $ $P_{an}$
  where \textit{true} assertion status is replaced by \textit{check}
  \State $P_{rtcheck} \leftarrow $ instrument $P_{check}$ with
    \textit{run-time checks}
  \Repeat
    \State Choose an exported
      predicate $p$ and generate a test case $input$ %
    \If{$p(input)$ in $P_{check}$ produces a run-time error at line $l$}
      \State $result \leftarrow$ \textsc{Error}$(input,l)$
    \ElsIf{maximum time or number of test executions is reached}
      \State $result \leftarrow$ \textsc{Timeout}
    \EndIf
  \Until{$result \neq$ \textsc{None}}
  \Return{$result$}
\EndProcedure

\end{algorithmic}
\end{algorithm}

We now show the concrete algorithm for implementing our proposal,
i.e., the driver that combines and inter-operates the different
components of the framework to achieve the desired results.
The essence of the algorithm (Alg.~\ref{alg:anatest}) is the following:
non-deterministically choose a program $P$ and a domain $\mathcal{D}$ from a
collection of benchmarks and domains, and execute the
\textsc{AnaTest}($P,\mathcal{D}$) procedure until an error is found or a limit
is reached.
Unless the testing part is ensured to explore the complete execution
space, it could in principle be useful to revisit the same
$(P,\mathcal{D})$ pair more than once.%
\footnote{Clearly, coverage of the program and coverage of the
  analyzer code could be a useful metric here to decide when to
  finish.}
There is no restriction regarding the number of entry
points or inputs to a program to be analyzed for. 
It is common in tools related to ours to use as benchmark programs
with a single entry point with no inputs (e.g., just a single
\texttt{void main()} function as entry point for C).
In \ciao program signatures and types
are optional.  Admissible inputs (i.e., the initial set of
possible states for analysis or test case generation) can be specified
by writing assertions for the exported predicates
or skipped altogether.
Note also that if the program $P$ had the restriction mentioned above
(in our case, exporting only a \texttt{main/0} predicate), then test
case generation would not be needed for our algorithm.
In the absence of %
assertions, the test case generation
framework has already some mechanisms
\dfnote{TODO: Which mechanism?}
to generate relevant test cases,
instead of random, nonsensical inputs which would exercise few
run-time checks before failing.
However, these generators have limitations, and the assertion-based testing
framework is in fact best used with assertions that have
descriptive-enough calling contexts,
or with custom user-defined
generators in their absence.

When the algorithm detects a faulty program-point assertion for some
$input$ (\textsc{Error}($input,l$)), it means that the concrete
execution reaches a state not captured by the (potentially safely
over-approximated) result of analysis.
It is important to note that although error diagnosis and debugging are
primarily left for the user to manually perform, our tool facilitates
the task in some aspects.
Firstly, it is possible to reconstruct (or store together with the test
output) additional information 
comparing the concrete execution trace (which is logged during testing) with
the analysis graph (recoverable from $P_{an}$, the program annotated with analysis results),
domain
operations (inspecting the analysis graph), and transfer functions
(from predicates that are \emph{native} to each
domain).\jfnote{automatically doing it is out of the scope}
Secondly, the \textit{assertion-based testing} tool supports shrinking
of failed test cases, so we can expect reasonably small variable
substitutions in the errors reported.
\dfnote{Commented out reference to delta debugging}
Lastly, as sketched in Algorithm~\ref{alg:anatest},
the error location and trace reported by the run-time verification framework
provide an approximate idea of the point where the
analysis went wrong, even if not necessarily of the original reason
why (which requires a different step of diagnosis).
\mhnote{Self-reminder: it is more complicated than this. DF: I agree!
  We should comment it out. It could be an idea for a paper
  (trace back, e.g. with backwards analysis, and identify
  the initial point of failure).}
If the run-time check error points to a
program-point assertion right after a call to an imported
predicate, then the analysis erred in applying the \textit{entry} declaration
for the predicate, the \textit{entry} declaration was wrong,
or if there was no \textit{entry} declaration, the analysis failed to
compute the ``topmost'' abstract state reachable from the call abstract
state.

\subsection{Some Considerations on Properties}
\label{sec:properties}

In order to test an
analysis with the algorithm proposed,
two conditions must be met. The first one is to have a translation
from the internal representation of the abstract values in the domain
to \ciao user-level properties.
These properties that can represent the information inferred by a
given domain or domains are called the \emph{native properties} of
the domain(s).
Note that these 
are already requirements for any abstract domain intended to make full
use of the framework, so normally all implemented domains include the
definition of the corresponding native properties and the translation
from abstract domain values to them.

The other condition is to be able to perform run-time checks for those
properties, i.e., that they can be used by the run-time checking
framework.
As discussed in Section~\ref{sec:preliminaries}, 
such run-time checks can range from very simple or even already built
into the language (like, e.g., var/1), to intermediate (like, e.g.,
aliasing or groundness), to more complex and costly (like, e.g., costs
or side-effects), to theoretically impossible (like, e.g.,
termination). But they can also be safely approximated to detect
errors or to issue warnings (e.g., in the case of termination by
detecting repeated identical calls which lead to non-termination or by
timeouts).  It is also important to note that complex analyses such as
termination are typically dependent on a number of other instrumental
analyses which can themselves be checked.

In general, the availability of run-time checks is a standard
requirement for domains to be able to make full use of the framework,
in order to support the dynamic checks that are generated when
properties cannot be proved statically. This functionality is normally
implemented when a new abstract domain is added to the system, by also
defining the related properties to be used in assertions.
If the definition of these native properties is provided directly in the
source language, then such properties are typically already runnable
and thus available for run-time checking; however, it is also possible
to provide an implementation specialized for run-time checking if
desired.
For properties that are declared native but are not written in the
source language, then a run-time test version must be provided.
In practice, most current \ciao abstract domains include the
mentioned functionalities and can be tested as is with the proposed
approach.

\subsection{Multivariance and Path-Sensitivity}
\label{sec:multivariance}

As presented, it could appear that our approach could miss some analysis
errors even if the right test cases are used, since we have, to all
appearances, disregarded \emph{multi-variance and path-sensitivity}.
In fact, in \ciaopp the information inferred is fully multi-variant,
and separate path information is kept for each variant (i.e., calling context). However, by
default, the analyzer produces an output that is easy for the programmer to inspect,
i.e., close to the source program.
This means that when outputting the analysis
results, by default the different versions of each
predicate (and the associated information) are combined into a single code version
and a single combined assertion for each program point and predicate.
If this default output is used when implementing our approach, it is
indeed entirely possible that the analysis errs at a program point
in one path but the algorithm never detects it: 
this can happen if, for example, in another path leading to the same
program point (such that the two paths and their corresponding analysis
results are collapsed --lubbed-- together at the same program point) the
analysis infers a too general value (higher in the domain lattice) at
that program point and thus, the error is not detected.
This issue is controlled by a flag that, when enabled, 
ensures that the different \emph{versions} are
not collapsed and are instead \emph{materialized} into different
predicate instances.
This way, multiple \emph{versions} may be generated for a given predicate, if
there are separate paths to them with different abstract states,
and the corresponding analysis information will be annotated
separately for each abstract path through the program in the program
text of the different versions, avoiding the problem mentioned above.

\section{Evaluation}
\label{evaluation}

In this section we report on the different experiments that we have
conducted in order to benchmark the checkification approach and assess
its practicality.

\subsection{Evaluation Setup}
\label{setup}

\paragraph{\textbf{The experiments.}}
\label{experiments}

The experiments have consisted in, for a set of benchmarks, analyzing
them with different abstract domains, performing the checkification
transformation on the analysis results, and testing the resulting
programs on sample inputs with run-time checks activated. 
The objective has been to assess whether we can indeed find errors
using the technique and to estimate the cost involved in detecting
those errors.

\paragraph{\textbf{Analyzer configuration.}}

The experiments were run with \ciao/\ciaopp version 1.23.  The tested
analyses used the standard, default configuration of the abstract
interpretation framework, i.e., the default values of the different
flags, such as, for example, using multi-variance on calls, using the
original \textit{PLAI} fixpoint algorithm, etc., but, of course, differ
in the abstract domain selected for performing each analysis.

\paragraph{\textbf{Properties and domains.}}

\begin{table}[!t]
  \centering
  \caption{Abstract Domains.}\label{domains}
  {\tablefont\begin{tabular}{@{\extracolsep{\fill}}cccc}
   \topline \textbf{Abstract} & \textbf{Properties} & & \\ \textbf{Domain} & \textbf{Abstracted} & \textbf{Maturity Level} & \textbf{Reference} \midline
\textit{gr} & aliasing, modes & intermediate & \cite{ciaopp-tutorial} \\
\textit{def} & aliasing, modes & intermediate & \cite{anconsall-acm} \\
\textit{sharing} & aliasing, modes & mature & \cite{ai-jlp} \\
\textit{shfr} & aliasing, modes & mature & \cite{freeness-iclp91} \\
\textit{shfr\texttt{+}nonvar} & aliasing, modes & intermediate & \\
\textit{shareson} & aliasing, modes & intermediate & \cite{comdom} \\
\textit{shfrson} & aliasing, modes & intermediate & \\   
\textit{son} & aliasing, modes & mature & \cite{sonder86} \\
\textit{share\_amgu} & aliasing, modes & mature & \\
\textit{shfr\_amgu} & aliasing, modes & mature & \\
\textit{shfrlin\_amgu} & aliasing, modes, linearity  & mature & \\
\textit{share\texttt{+}clique} & aliasing, modes & mature & \cite{shcliques-padl06} \\
\textit{shfr\texttt{+}clique} & aliasing, modes & mature & \cite{shcliques-padl06} \\
\textit{share\texttt{+}clique\texttt{+}def} & aliasing, modes & experimental & \\
\textit{shfr\texttt{+}clique\texttt{+}def}  & aliasing, modes & experimental & \\
\textit{eterms} & types & mature & \cite{eterms-sas02} \\
\textit{polyhedra} & numerical & experimental & \cite{BagnaraRZH02} \\
\textit{depth-k} & term structure & intermediate & \cite{sato-tamaki} \\
\textit{det} & determinacy & mature & \cite{determ-lopstr04,determinacy-ngc09} \\
\textit{nfg} & (non)failure & intermediate & \cite{non-failure-iclp97,nfplai-flops04-medium} 
  \botline
  \end{tabular}}
\end{table}

In the experiments we used a wide range of analyses for different
properties that are typically of interest when describing or
verifying %
logic programs (the list of all the abstract domains used is
provided in Table~\ref{domains}):\footnote{The table also provides
  references for each domain, except for some that are combinations of
  other domains not explicitly described in other papers.}
\begin{itemize}
\item 
The first class of properties is aimed at capturing variable
instantiation state, i.e., which variables 
are bound to ground terms, or free, and, if
they are not ground, the variable sharing relationships among them.
These properties are approximated using
\textit{aliasing} and \textit{modes}-style abstract domains.

\item 
The second set of properties refers to the
shapes of the data structures constructed by the program
in memory. These properties are tracked by the 
\textit{term structure} 
and 
\textit{types} classes of abstract domains.

\item The third class of properties that we have considered refers to the
numerical relations among program variables,
which are useful to describe properties of numerical parts of
programs. %
For these, we use in our experiments 
\textit{numerical}-style abstract domains.

\item Finally, we also evaluate the approach on analyses for computational
  properties, i.e., properties of whole computation subtrees, in
  particular  \textit{determinacy} and \textit{(non)failure}.
These analyses sometimes do not provide the information at
the program points between literals, but rather at the predicate
level. In these cases, the transformation and tests are done at
the predicate level. 
\end{itemize}

\begin{table}[!t]
  \centering
  \caption{Benchmarks.}\label{benchmarks}
  {\tablefont\begin{tabular}{@{\extracolsep{\fill}}ll}
    \topline \texttt{Bench} & \midline
  \texttt{mmatrix}   & matrix multiplication for two matrices with dimensions $n \times n$; \\
  \texttt{qsort}     & the quicksort program; \\
  \texttt{exp}       & exponential calculation; \\
  \texttt{aiakl}     & initialization for abstract unification in AKL analyzer; \\
  \texttt{ham}       & a program that generates the sequence of Hamming numbers; \\
  \texttt{fft}       & fast Fourier transformation calculation;  \\
  \texttt{factorial} & recursive factorial calculation; \\
  \texttt{witt}      & the WITT clustering system implementation; \\
  \texttt{poly}      & a program that raises a polynomial ($1 + x + y + z$) to the 10th power symbolically; \\
  \texttt{deriv}     & symbolic differentiation of a given formula; \\
  \texttt{grammar}   & a simple sentence parser;\\
  \texttt{fib}       & a program that finds $N$-th Fibonacci number; \\
  \texttt{boyer}     & a theorem prover implementation based on Lisp by R.\ Boyer (nqthm system),\\ & performs symbolic evaluation of a formula;  \\
  \texttt{queens}    & the $N$ queens program (number of the queens being the input); \\
  \texttt{jugs}      & the water jugs problem;\\
  \texttt{bid}       & compute opening bid for bridge hand;\\
  \texttt{nreverse}  & naive list reversal; \\
  \texttt{guardians} & prison guards game; \\
  \texttt{crypt}     & crypto-multiplication puzzle solver; 
  \botline
  \end{tabular}}
\end{table}

\paragraph{\textbf{Programs analyzed.}}
The programs used in our experiments can be divided into two different
groups: 

The first group, listed in Table~\ref{benchmarks},
comprises a number of well-known, classic benchmarks. Some of them 
also represent kernels of applications.  For example, \texttt{aiakl}
is the main part of an
analyzer for the AKL language; \texttt{boyer} is the kernel of a
theorem prover; and \texttt{witt} is
the central part of a conceptual clustering application.

The second group %
comprises %
complete systems that are in current use: 
\begin{itemize}
\item A \texttt{filtering} tool~\citep{hald-demo-iclp-tc}, used
  regularly for creating teaching materials.  This tool
  contains %
  1.1K lines of code in its kernel, and uses also a number of \ciao
  libraries.  It is 
  an interesting example since it includes different \textit{built-in}
  predicates for handling files and streams.

\item \texttt{Deepfind}~\citep{deepfind-iclp2016}, a tool that facilitates
  searching code repositories and libraries by querying for semantic
  characteristics of the code. Its kernel consists of around 10 files where
  the Prolog code is about 1.5K lines, and uses in addition a good number of
  \ciaopp libraries to perform program analysis and assertion
  comparison.

\item The classic \texttt{chat-80} program~\citep{chat}, a natural language
  interface to a geographical database. It comprises 4.8k lines of
  code across 22 files. While typically used more as a demo than a
  real application, we have included it in this group because it is of
  good size, contains a number of system libraries using different
  Prolog \textit{built-ins} and library predicates, and is known to
  stress several abstract interpretation domains.

\item \texttt{LPdoc}, %
a
  documenter for LP systems used by \ciao and
  XSB~\citep{lpdoc-entcs,lpdoc-reference}.  This is
  the largest example with its kernel Prolog code being analyzed comprising
  about 22K lines, plus the use of many \ciao libraries. 

\item \texttt{Spectector}~\citep{spectector}, a tool for automatically detecting
  leaks introduced by speculatively executed instructions in x64 assembly programs.
  It consists of 15 modules with around 1.6K lines of code.

\item The \texttt{s(CASP)}~\citep{scasp-iclp2018} system is a top-down interpreter for ASP
  programs with constraints. The Prolog code is distributed into 44 modules.

\end{itemize}
No %
special analysis-related criteria were used in benchmark
selection, and 
the code was analyzed and run as is, without modifications.
The classic benchmarks, by default, include annotations with program assertions that
describe the expected behavior, while no additional information is provided
in the real-world programs.

\newcommand{\msp}{\textcolor{white}{no}}
\newcommand{\errorlegend}{Classes of bugs found are marked with: \,
  \dbox{\msp}                   \, \texttt{=} abstract domain implementation, \,
  \dotuline{\msp}                  \texttt{=} fixpoint algorithms, \, 
  \tbox{\msp}                   \, \texttt{=} semantic inconsistencies, \,
  \uwave{\msp}                     \texttt{=} run-time check instrumentation, \,
  \underline{\underline{\msp}}     \texttt{=} third-party libraries.}

\begin{table}[!t]
  \centering
  \caption{Timings and errors detected (1).}\label{benchmarkstimes1}
  {\tablefont\begin{tabular}{@{\extracolsep{\fill}}lcccccccccc}
    \topline
    \multicolumn{1}{c}{} & \multicolumn{10}{c}{\bf{Absolute run time, s (Analysis time, s)}}  \\
    \cmidrule(rl){2-11} 
    \multicolumn{1}{c}{Program} & \multicolumn{2}{c}{\textit{gr}} & \multicolumn{2}{c}{\textit{def}} & \multicolumn{2}{c}{\textit{sharing}} & \multicolumn{2}{c}{\textit{sharefree}} & \multicolumn{2}{c}{\textit{shfr\texttt{+}nonvar}} \\
    \cmidrule(rl){2-3} \cmidrule(rl){4-5}  \cmidrule(rl){6-7}  \cmidrule(rl){8-9}  \cmidrule(rl){10-11}
    \texttt{mmatrix}   & 0.2                        & (0.25)   & 0.2                        & (0.25)      & 0.4                         & (0.25)    &  0.4                        & (0.25)       & 0.4             & (0.25)   \\[0.07cm]
    \texttt{qsort}     & 1.0                        & (0.21)   & 1.0                        & (0.21)      & 1.7                         & (0.21)    &  1.7                        & (0.22)       & 1.7             & (0.22)   \\[0.07cm]
    \texttt{exp}       & 0.4$\cdot$$10^{\texttt{-}1}$ & (0.22)   & 0.4$\cdot$$10^{\texttt{-}1}$ & (0.22)      & 0.2                         & (0.23)    &  0.2                        & (0.23)       & 0.2             & (0.23)   \\[0.07cm]
    \texttt{aiakl}     & 0.2$\cdot$$10^{\texttt{-}3}$ & (0.25)   & 0.3$\cdot$$10^{\texttt{-}3}$ & (0.24)      & 0.8$\cdot$$10^{\texttt{-}3}$ & (0.25)     &  0.9$\cdot$$10^{\texttt{-}3}$ & (0.26)       & 0.9$\cdot$$10^{\texttt{-}3}$ & (0.25)   \\[0.07cm]
    \texttt{ham}       & 0.1                        & (0.33)   & 0.1                        & (0.34)      & 0.3                         & (0.34)     &  0.3                        & (0.34)       & \dbox{n\texttt{\texttt{/}}a$^{\dag}$}      & (0.26)   \\[0.07cm]
    \texttt{fft}       & 0.9                        & (0.36)   & \tbox{$2.5^{\dag}$}                 & (0.35)      & 6.8                         & (0.36)     &  6.8                        & (0.36)       & 6.8             & (0.37)   \\[0.07cm]
    \texttt{factorial} & 0.1                        & (0.20)   & 0.1                        & (0.20)      & 0.2                         & (0.20)     &  0.2                        & (0.21)       & 0.2             & (0.20)   \\[0.07cm]
    \texttt{witt}      & 11.4                       & (1.08) & 23.2                         & (1.08)    & 57.7                          & (1.12)   &  60.9                         & (1.14)     & \dbox{n\texttt{\texttt{/}}a$^{*}$}      & (1.14) \\[0.07cm]
    \texttt{poly}      & 0.8                        & (0.31)   & 8.7                        & (0.34)      & 32.3                        & (0.44)     &  30.9                       & (0.42)       & 31.7            & (0.52)   \\[0.07cm]
    \texttt{deriv}     & 0.2$\cdot$$10^{\texttt{-}3}$ & (0.28)   & 0.5$\cdot$$10^{\texttt{-}3}$ & (0.28)      & 2.4$\cdot$$10^{\texttt{-}3}$ & (0.29)     &  0.2$\cdot$$10^{\texttt{-}2}$ & (0.29)       & 0.3$\cdot$$10^{\texttt{-}2}$ & (0.29)   \\[0.07cm]
    \texttt{grammar}   & 0.2$\cdot$$10^{\texttt{-}4}$ & (0.23)   & 0.5$\cdot$$10^{\texttt{-}4}$ & (0.22)      & 1.5$\cdot$$10^{\texttt{-}4}$ & (0.23)     &  0.2$\cdot$$10^{\texttt{-}3}$ & (0.23)       & 0.2$\cdot$$10^{\texttt{-}3}$ & (0.24)   \\[0.07cm] 
    \texttt{fib}       & 0.1                        & (0.21)   & 0.1                        & (0.21)      & 0.3                         & (0.20)    &  0.3                         & (0.20)       & 0.3             & (0.29)   \\[0.07cm]
    \texttt{boyer}     & 0.7                        & (0.50)   & 2.7                        & (0.52)      & 15.8                        & (0.56)    &  16.0                        & (0.56)       & 16.2            & (0.57)   \\[0.07cm]
    \texttt{queens}    & 6.0                        & (0.23)   & 6.0                        & (0.24)      & 12.4                        & (0.23)    &  13.2                        & (0.24)       & 13.2            & (0.23)   \\[0.07cm]
    \texttt{jugs}      & 0.2$\cdot$$10^{\texttt{-}3}$ & (0.31)   & 0.9$\cdot$$10^{\texttt{-}3}$ & (0.32)      & 0.3$\cdot$$10^{\texttt{-}2}$  & (0.34)    &  0.3$\cdot$$10^{\texttt{-}2}$ & (0.34)       & 0.3$\cdot$$10^{\texttt{-}2}$ & (0.32)   \\[0.07cm]
    \texttt{bid}       & 0.2$\cdot$$10^{\texttt{-}2}$ & (0.38)   & 0.2$\cdot$$10^{\texttt{-}2}$ & (0.37)      & 0.4$\cdot$$10^{\texttt{-}2}$  & (0.38)    &  0.5$\cdot$$10^{\texttt{-}2}$ & (0.38)       & \dbox{n\texttt{\texttt{/}}a$^{*}$}       & (0.39)   \\[0.07cm]
    \texttt{nreverse}  & 0.3$\cdot$$10^{\texttt{-}1}$ & (0.20)   & 1.3                        & (0.21)      & 4.0                         & (0.22)    &  \uwave{4.0}                & (0.22)       & 4.1             & (0.21)   \\[0.07cm]
    \texttt{guardians} & 0.4                        & (0.28)   & 0.4                        & (0.28)      & 0.7                         & (0.29)    &  0.7                        & (0.29)       & \dbox{n\texttt{\texttt{/}}a$^{*}$}       & (0.29)   \\[0.07cm]
    \texttt{crypt}     & 0.1$\cdot$$10^{\texttt{-}1}$ & (0.31)   & 0.1$\cdot$$10^{\texttt{-}1}$ & (0.31)      & n\texttt{/}a               & (n\texttt{/}a)     &  0.1                        & (0.32)       & 0.1             & (0.31) \\
    \hline
    \texttt{exfilter} & 0.2 & (6.10) & 21.9 & (6.84)  & 83.4 & (14.58) & 83.6  & (13.55)  & n\texttt{/}a & (n\texttt{/}a) \\[0.07cm]
    \texttt{deepfind} & 0.4 & (7.12) & 0.3 & (6.39) & 1.9 & (23.46) & 2.1 & (14.06) & 1.0 & (9.72) \\[0.07cm]
    \texttt{chat-80} & 0.4$\cdot$$10^{\texttt{-}1}$ & (5.41) & \dbox{n\texttt{/}a$^{*}$} & (5.78) & 5.4 & (54.60) & 5.5 & (54.28) & 5.7 & (54.29) \\[0.07cm]
    \texttt{LPdoc} & 0.3 & (22.20) & 1.0 & (23.03) & 52.3 & (72.45)  & 53.6 & (65.06) & 53.0 & (112.12) \\[0.07cm]
    \texttt{Spectector} & 0.6 & (4.27) & \dbox{n\texttt{/}a$^{*}$} & (4.43) & 29.1 & (5.72) & 24.0 & (5.75) & 24.5  & (5.77) \\[0.07cm]
    \texttt{s(CASP)} & 0.6 & (10.35) & 24.2 & (10.69) & 56.1 & (237.75) & 56.6 & (197.62) & \dbox{n\texttt{/}a$^{*}$} & (205.86)
     \botline
  \end{tabular}}
\begin{tablenotes}
\item \errorlegend
\end{tablenotes}
\end{table}

\begin{table}[!t]
  \centering
  \caption{Timings and errors detected (2).}\label{benchmarkstimes2}
  {\tablefont\begin{tabular}{@{\extracolsep{\fill}}lcccccccccc}
    \topline
    \multicolumn{1}{c}{} & \multicolumn{10}{c}{\bf{Absolute run time, s (Analysis time, s)}}\\
    \cmidrule(rl){2-11} 
    \multicolumn{1}{c}{Program} & \multicolumn{2}{c}{\textit{shareson}} & \multicolumn{2}{c}{\textit{shfrson}} & \multicolumn{2}{c}{\textit{sondergaard}} & \multicolumn{2}{c}{\textit{share\_amgu}} & \multicolumn{2}{c}{\textit{shfr\_amgu}}\\
    \cmidrule(rl){2-3} \cmidrule(rl){4-5}  \cmidrule(rl){6-7}  \cmidrule(rl){8-9}  \cmidrule(rl){10-11} 
    \texttt{mmatrix}   & 0.6                        & (0.25)     & 0.7                        & (0.25)  & 0.4                        & (0.25)  & 0.4                         & (0.25)   & 0.4                          & (0.26)       \\[0.07cm]
    \texttt{qsort}     & 2.8                        & (0.21)     & 2.8                        & (0.21)  & 1.7                        & (0.21)  & 1.7                         & (0.21)   & 1.7                          & (0.21)       \\[0.07cm]
    \texttt{exp}       & 0.2                        & (0.23)     & 0.2                        & (0.23)  & 0.2                        & (0.23)  & 0.2                         & (0.23)   & 0.2                          & (0.23)       \\[0.07cm]
    \texttt{aiakl}     & 0.1$\cdot$$10^{\texttt{-}2}$ & (0.26)     & 0.1$\cdot$$10^{\texttt{-}2}$ & (0.26)  & 0.1$\cdot$$10^{\texttt{-}2}$ & (0.26)   & 0.8$\cdot$$10^{\texttt{-}3}$ & (0.25)    & 0.9$\cdot$$10^{\texttt{-}3}$ & (0.26)       \\[0.07cm]
    \texttt{ham}       & 0.5                        & (0.36)     & 0.5                        & (0.37)  & 0.4                        & (0.34)  & 0.3                         & (0.35)   & 0.3                          & (0.34)       \\[0.07cm]
    \texttt{fft}       & 9.5                        & (0.38)     & 9.6                        & (0.38)  & 7.5                        & (0.36)  & 6.8                         & (0.36)   & 6.8                          & (0.37)       \\[0.07cm]
    \texttt{factorial} & 0.3                        & (0.20)     & 0.3                        & (0.21)  & 0.2                        & (0.20)  & 0.2                         & (0.20)   & 0.2                          & (0.20)       \\[0.07cm]
    \texttt{witt}      & 60.7                       & (3.93)     & n\texttt{/}a               & (n\texttt{/}a) & n\texttt{/}a        & (n\texttt{/}a) & 57.7                 & (1.10) & 61.0                         & (1.14)     \\[0.07cm] 
    \texttt{poly}      & 31.8                       & (0.48)     & 32.0                       & (0.49)  & 31.7                       & (0.39)  & 32.4                        & (0.57)   & 31.2                         & (0.48)       \\[0.07cm]
    \texttt{deriv}     & 0.3$\cdot$$10^{\texttt{-}2}$ & (0.29)     & 0.4$\cdot$$10^{\texttt{-}2}$ & (0.30)  & 0.2$\cdot$$10^{\texttt{-}2}$ & (0.29) & 0.2$\cdot$$10^{\texttt{-}2}$  & (0.29)   & 0.2$\cdot$$10^{\texttt{-}2}$   & (0.29)       \\[0.07cm]
    \texttt{grammar}   & 0.2$\cdot$$10^{\texttt{-}3}$ & (0.23)     & 0.2$\cdot$$10^{\texttt{-}2}$ & (0.24)  &  0.2$\cdot$$10^{\texttt{-}3}$ & (0.24) & 0.1$\cdot$$10^{\texttt{-}3}$  & (0.24)   & 0.2$\cdot$$10^{\texttt{-}3}$   & (0.23)       \\[0.07cm]
    \texttt{fib}       & 0.4                        & (0.21)     & 0.4                        & (0.21)  & 0.3                        & (0.20)  & 0.3                         & (0.22)   & 0.3                          & (0.21)       \\[0.07cm]
    \texttt{boyer}     & 20.3                       & (0.57)     & \tbox{20.7$^{\dag}$}                & (0.59)  & 19.7                       & (0.56)  & 15.7                        & (0.58)   & 16.2                         & (0.57)     \\[0.07cm]
    \texttt{queens}    & 21.6                       & (0.24)     & 23.1                       & (0.28)  & 13.6                       & (0.24)  & 12.3                        & (0.23)   & 13.4                         & (0.24)       \\[0.07cm]
    \texttt{jugs}      & 0.3$\cdot$$10^{\texttt{-}2}$ & (0.33)     & 0.3$\cdot$$10^{\texttt{-}2}$ & (0.33)  & 0.3$\cdot$$10^{\texttt{-}2}$ & (0.33) & 0.3$\cdot$$10^{\texttt{-}2}$  & (0.33)   & 0.3$\cdot$$10^{\texttt{-}2}$   & (0.33)       \\[0.07cm]
    \texttt{bid}       & 0.7$\cdot$$10^{\texttt{-}2}$ & (0.39)     & 0.7$\cdot$$10^{\texttt{-}2}$ & (0.40)  & 0.4$\cdot$$10^{\texttt{-}2}$ & (0.39) & 0.4$\cdot$$10^{\texttt{-}2}$  & (0.38)   & 0.4$\cdot$$10^{\texttt{-}2}$   & (0.39)       \\[0.07cm]
    \texttt{nreverse}  & 4.0                        & (0.22)     & 4.0                        & (0.22)  & 3.9                        & (0.22)  & 4.0                         & (0.21)   & 3.9                          & (0.22)       \\[0.07cm]
    \texttt{guardians} & 54.1                       & (0.29)     & 1.1                        & (0.31)  & 0.7                        & (0.29)  & 0.6                         & (0.29)   & 0.7                          & (0.29)       \\[0.07cm]
    \texttt{crypt}     & 0.1                        & (0.42)    & 0.1                        & (0.35)  & 0.1                        & (0.32)  & n\texttt{/}a                & (n\texttt{/}a)    & 0.1                          & (0.31)         
    \\
    \hline
    \texttt{exfilter} & n\texttt{/}a & (n\texttt{/}a) & n\texttt{/}a & (n\texttt{/}a)  & n\texttt{/}a & (54.82) & 83.7 & (14.91)  & 83.5 & (13.45) \\[0.07cm]
    \texttt{deepfind} & 1.9 & (9.49) & 2.0 & (10.12) & 1.6 & (50.12) & 0.5 & (49.88) & 0.4 & (12.14)  \\[0.07cm]
    \texttt{chat-80} & 0.4 & (56.53) & 0.5 & (55.72) & 0.5 & (59.15) & 0.4 & (43.86) & 0.5 & (39.94)\\[0.07cm]
    \texttt{LPdoc} & 53.0 & (160.86) & 53.5 & (119.31) & 52.1 & (22.89) & 52.7  & (200.87) & 52.5 & (262.71) \\[0.07cm]
    \texttt{Spectector} & 24.1 & (5.96) & 24.5 & (5.80) & 24.6 & (8.71) & 36.9 & (10.97) & 24.2 & (5.91)  \\[0.07cm]
    \texttt{s(CASP)} & 56.1 & (207.61) & 56.1 & (198.59) & 56.3 & (33.28) & 57.0 & (291.39) & 55.8 & (170.38) 
\botline
  \end{tabular}}
\begin{tablenotes}
\item \errorlegend 
\end{tablenotes}
\end{table}

\begin{table}[!t]
  \centering
  \caption{Timings and errors detected (3).}\label{benchmarkstimes3}
  {\tablefont\begin{tabular}{@{\extracolsep{\fill}}lcccccccccccc}
    \topline
    \multicolumn{1}{c}{} & \multicolumn{10}{c}{\bf{Absolute run time, s (Analysis time, s)}}\\
    \cmidrule(rl){2-11} 
    \multicolumn{1}{c}{Program} & \multicolumn{2}{c}{\textit{shfrlin\_amgu}} & \multicolumn{2}{c}{\textit{share\texttt{+}clique}} & \multicolumn{2}{c}{\textit{shfr\texttt{+}clique}} & \multicolumn{2}{c}{\textit{share\texttt{+}clique\texttt{+}def}} & \multicolumn{2}{c}{\textit{shfr\texttt{+}clique\texttt{+}def}}\\
    \cmidrule(rl){2-3} \cmidrule(rl){4-5}  \cmidrule(rl){6-7}  \cmidrule(rl){8-9}  \cmidrule(rl){10-11} 
    \texttt{mmatrix}   & 0.5                        & (0.26)      & 0.4                         & (0.25)      & 0.4                         & (0.25)   & 0.4                        & (0.26)   & 0.4                                & (0.25)    \\[0.07cm]
    \texttt{qsort}     & 1.8                        & (0.21)      & 1.7                         & (0.21)      & 1.7                         & (0.22)   & 1.7                        & (0.21)   & 1.7                                & (0.21)    \\[0.07cm]
    \texttt{exp}       & 0.2                        & (0.23)      & 0.4$\cdot$$10^{\texttt{-}1}$  & (0.23)      & 0.4$\cdot$$10^{\texttt{-}1}$  & (0.23)   & 0.4$\cdot$$10^{\texttt{-}1}$ & (0.22)   & \dbox{0.5$\cdot$$10^{\texttt{-}1}$$^{\dag}$} & (0.23)    \\[0.07cm]
    \texttt{aiakl}     & 0.1$\cdot$$10^{\texttt{-}2}$ & (0.26)      & 0.8$\cdot$$10^{\texttt{-}3}$  & (0.25)      & 0.9$\cdot$$10^{\texttt{-}3}$  & (0.26)   & 0.8$\cdot$$10^{\texttt{-}3}$ & (0.25)   & 0.9$\cdot$$10^{\texttt{-}3}$        & (0.25)    \\[0.07cm]
    \texttt{ham}       & 0.4                        & (0.35)      & 0.2                         & (0.34)      & 0.2                         & (0.35)   & 0.2                        & (0.36)    & \dbox{0.2$^{\dag}$}                        & (0.35)    \\[0.07cm]
    \texttt{fft}       & 7.1                        & (0.37)      & 6.7                         & (0.36)      & 6.8                         & (0.37)   & 6.8                        & (0.37)    & 6.8                               & (0.37)    \\[0.07cm]
    \texttt{factorial} & 0.2                        & (0.20)      & 0.2                         & (0.20)      & 0.2                         & (0.20)   & 0.2                        & (0.20)    & \dbox{0.2$^{\dag}$}                        & (0.20)    \\[0.07cm]
    \texttt{witt}      & 67.9                       & (1.18)      & \tbox{49.4$^{\dag}$}                 & (1.12)      & \dbox{n\texttt{/}a$^{\dag}$}         & (1.15) & \tbox{49.2$^{\dag}$}                  & (1.13)  & \dbox{n\texttt{/}a$^{*}$}               & (1.15)  \\[0.07cm] 
    \texttt{poly}      & 31.5                       & (0.53)      & 2.4                         & (0.36)      & \uwave{2.4$^{*}$}                 & (0.35)   & 2.4                        & (0.38)    & 2.4                               & (0.38)    \\[0.07cm]
    \texttt{deriv}     & 0.3$\cdot$$10^{\texttt{-}2}$ & (0.30)      & 0.2$\cdot$$10^{\texttt{-}2}$  & (0.29)      & 0.2$\cdot$$10^{\texttt{-}2}$  & (0.29)  & 0.2$\cdot$$10^{\texttt{-}2}$ & (0.29)    & \dbox{0.2$\cdot$$10^{\texttt{-}2}$$^{\dag}$} & (0.29)    \\[0.07cm]
    \texttt{grammar}   & 0.2$\cdot$$10^{\texttt{-}3}$ & (0.23)      & 0.2$\cdot$$10^{\texttt{-}3}$  & (0.23)      & 0.2$\cdot$$10^{\texttt{-}3}$  & (0.23)  & 0.2$\cdot$$10^{\texttt{-}3}$ & (0.24)    & 0.2$\cdot$$10^{\texttt{-}3}$        & (0.23)    \\[0.07cm]
    \texttt{fib}       & 0.3                        & (0.22)      & 0.3                         & (0.21)      & 0.3                         & (0.21)   & 0.3                        & (0.21)    & 0.3         & (0.21)    \\[0.07cm]
    \texttt{boyer}     & 17.1                       & (0.60)      & 15.1                        & (0.55)      & 15.2                        & (0.55)   & 15.0                       & (0.57)    & \dbox{15.5$^{\dag}$} & (0.59)    \\[0.07cm]
    \texttt{queens}    & 14.6                       & (0.25)      & 12.7                        & (0.24)      & 13.2                        & (0.23)   & 12.4                       & (0.24)    & 13.2        & (0.24)    \\[0.07cm]
    \texttt{jugs}      & 0.4$\cdot$$10^{\texttt{-}2}$ & (0.34)      & 0.6$\cdot$$10^{\texttt{-}3}$  & (0.32)      & 0.6$\cdot$$10^{\texttt{-}3}$  & (0.32)  & 0.6$\cdot$$10^{\texttt{-}3}$ & (0.32)    & \dbox{0.6$\cdot$$10^{\texttt{-}3}$$^{\dag}$} & (0.32)    \\[0.07cm]
    \texttt{bid}       & 0.5$\cdot$$10^{\texttt{-}2}$ & (0.42)      & 0.4$\cdot$$10^{\texttt{-}2}$  & (0.39)      & \dbox{n\texttt{/}a$^{\dag}$}         & (0.39)  & 0.3$\cdot$$10^{\texttt{-}2}$ & (0.39)    & \dbox{n\texttt{/}a$^{*}$}               & (0.39)    \\[0.07cm]
    \texttt{nreverse}  & 4.1                        & (0.21)      & 0.1                         & (0.22)      & 0.1                         & (0.21)   & 0.1                        & (0.21)    & 0.2                               & (0.21)    \\[0.07cm]
    \texttt{guardians} & 0.7                        & (0.29)      & 0.7                         & (0.29)      & \dbox{n\texttt{/}a$^{\dag}$}         & (0.30)   & 0.7                       & (0.29)    & \dbox{n\texttt{/}a$^{*}$}                & (0.23)    \\[0.07cm]
    \texttt{crypt}     & 0.1                        & (0.33)      & 0.9$\cdot$$10^{\texttt{-}2}$  & (71.83)     & 0.1                         & (0.32)   & 0.1$\cdot$$10^{\texttt{-}1}$ & (71.32) & \dbox{0.8$\cdot$$10^{\texttt{-}1}$$^{\dag}$} & (0.32)
    \\
    \hline
    \texttt{exfilter} & 87.1 & (15.31) & 85.4 & (20.72)  & 86.3 & (18.59) & 84.5 & (20.40)  &  \dbox{n\texttt{/}a$^{*}$}  & (15.25) \\[0.07cm]
    \texttt{deepfind} & 1.5 & (15.69) & 1.8 & (419.87) & 2.1 & (28.18) & 1.0 & (3.19) & \dbox{n\texttt{/}a$^{*}$} & (3.20) \\[0.07cm]
    \texttt{chat-80} & 0.6 & (60.55) & 0.2 & (5.77) & 0.2 & (5.82) & 0.1 & (5.92) & \dbox{n\texttt{/}a$^{*}$} & (5.39) \\[0.07cm]
    \texttt{LPdoc} & 51.6 & (300.92) & 0.7 & (73.90)  & 0.8 & (179.85) & 0.7  & (74.94) & \dbox{n\texttt{/}a$^{*}$}  & (24.83) \\[0.07cm]
    \texttt{Spectector} & 24.7 & (6.02) & 3.7 & (4.35) & 3.8 & (4.69) & 1.7 & (4.32) & \dbox{n\texttt{/}a$^{*}$} & (3.28)  \\[0.07cm]
    \texttt{s(CASP)} & 56.0 & (328.94) & 10.1 & (25.94) & \dbox{n\texttt{/}a$^{\dag}$} & (24.37) & 8.0 & (68.63) & \dbox{n\texttt{/}a$^{*}$} & (50.10)
\botline 
  \end{tabular}}
\begin{tablenotes}
\item \errorlegend 
\end{tablenotes}
\end{table}

\begin{table}[!t]
  \centering
  \caption{Timings and errors detected (4).}\label{benchmarkstimes4}
  {\tablefont\begin{tabular}{@{\extracolsep{\fill}}lcccccccccccc}
    \topline
    \multicolumn{1}{c}{} & \multicolumn{10}{c}{\bf{Absolute run time, s (Analysis time, s)}}\\
    \cmidrule(rl){2-11} 
    \multicolumn{1}{c}{Program} & \multicolumn{2}{c}{\textit{eterms}} & \multicolumn{2}{c}{\textit{polyhedra}} & \multicolumn{2}{c}{\textit{depth-k}} & \multicolumn{2}{c}{\textit{det}} & \multicolumn{2}{c}{\textit{nfg}} \\
    \cmidrule(rl){2-3} \cmidrule(rl){4-5}  \cmidrule(rl){6-7}  \cmidrule(rl){8-9} \cmidrule(rl){10-11} 
    \texttt{mmatrix}    & 0.6                                                    & (0.64)        & 0.2$\cdot$$10^{\texttt{-}1}$            & (0.25)          & \uwave{0.1$\cdot$$10^{\texttt{-}1}$$^{\dag}$}              & (0.26)         & 0.3                                 & (0.40)   & 0.3 & (0.37) \\[0.09cm]
    \texttt{qsort}      & 4.0                                                    & (0.53)        & n\texttt{/}a                          & (n\texttt{/}a)  & \uwave{0.2$\cdot$$10^{\texttt{-}1}$$^{\dag}$}              & (0.21)         & 4.4                                 & (0.47)   & 4.4 & (0.46) \\[0.09cm] 
    \texttt{exp}        & 0.1                                                    & (0.27)        & 0.4$\cdot$$10^{\texttt{-}1}$            & (0.25)          & 0.2$\cdot$$10^{\texttt{-}1}$                      & (0.22)         & 0.2                                 & (0.20)   & 0.2 & (0.23)  \\[0.09cm] 
    \texttt{aiakl}      & \underline{\underline{0.8$\cdot$$10^{\texttt{-}3}$$^{\dag}$}}     & (0.48)        & \dbox{n\texttt{\texttt{/}}a$^{\dag}$}          & (0.23)          & 0.3$\cdot$$10^{\texttt{-}4}$                      & (0.25)         & 0.1$\cdot$$10^{\texttt{-}3}$          & (0.51)   & 0.1$\cdot$$10^{\texttt{-}2}$      & (0.39)     \\[0.09cm]
    \texttt{ham}        & 7.6                                                    & (0.48)        & \dbox{n\texttt{\texttt{/}}a$^{\dag}$}          & (0.35)          & \uwave{0.5$\cdot$$10^{\texttt{-}1}$$^{\dag}$}              & (0.33)         & 0.8$\cdot$$10^{\texttt{-}1}$          & (0.35)    & 0.8$\cdot$$10^{\texttt{-}1}$  & (0.27) \\[0.09cm] 
    \texttt{fft}        & \underline{\underline{9.8}$^{\dag}$}                            & (3.00)        & 0.5$\cdot$$10^{\texttt{-}2}$            & (0.25)          & \uwave{0.3$^{\dag}$}                                     & (0.32)         &  n\texttt{/}a                       & (4.34)     & 13.2    & (3.46) \\[0.09cm]   
    \texttt{factorial}  & 0.2                                                    & (0.22)        & 0.7$\cdot$$10^{\texttt{-}1}$            & (0.21)          &  0.2$\cdot$$10^{\texttt{-}1}$                     & (0.17)         & 0.2                                 & (0.18)   & 0.3 & (0.19) \\[0.09cm]
    \texttt{witt}       & \dbox{n\texttt{/}a$^{*}$}                                    & (0.48)        & n\texttt{/}a                          & (n\texttt{/}a)  & n\texttt{/}a                                    & (n\texttt{/}a) & \dbox{n\texttt{/}a$^{\dag}$}                  & (0.35)    & n\texttt{/}a & (n\texttt{/}a) \\[0.09cm] 
    \texttt{poly}       & \underline{\underline{27.8}$^{\dag}$}                           & (15.68)       & 0.2                                   & (0.25)          & 0.5                                             & (0.33)         & \dbox{n\texttt{/}a$^{\dag}$} & (32.70) & n\texttt{/}a & (n\texttt{/}a) \\[0.09cm]
    \texttt{deriv}      & 0.2                                                    & (2,384.47)    & 0.2$\cdot$$10^{\texttt{-}3}$            & (0.29)          & 0.1$\cdot$$10^{\texttt{-}3}$                      & (0.30)         & n\texttt{\texttt{/}}a        & n\texttt{\texttt{/}}a &  n\texttt{/}a & (n\texttt{/}a)    \\[0.09cm]         
    \texttt{grammar}    & 0.1$\cdot$$10^{\texttt{-}3}$                             & (0.37)        & n\texttt{/}a                          & (n\texttt{/}a)  & \uwave{0.4$\cdot$$10^{\texttt{-}4}$$^{\dag}$}              & (0.21)         & 0.1$\cdot$$10^{\texttt{-}3}$          & (0.26)  &  0.2$\cdot$$10^{\texttt{-}3}$  & (0.27) \\[0.09cm] 
    \texttt{fib}        & 0.2                                                    & (0.27)        & 0.1                                   & (0.23)          & 0.1$\cdot$$10^{\texttt{-}1}$                      & (0.20)         & 0.3                                 & (0.19)  & 0.6 & (0.20)  \\[0.09cm] 
    \texttt{boyer}      & 9.0                                       & (6.41)    & \dbox{n\texttt{\texttt{/}}a$^{\dag}$}  & (0.37)  &  0.4        & (0.56)           & 55.0                                             & (7.98)  & n\texttt{/}a & (n\texttt{/}a)  \\[0.09cm]   
    \texttt{queens}     & 21.1                                                   & (0.36)        & 1.3                                   & (0.25)          & \uwave{1.0$^{\dag}$}                                     & (0.23)         & 26.7                                & (0.26)  & 27.3 & (0.26) \\[0.09cm]
    \texttt{jugs}       & \underline{\underline{1.6}$^{\dag}$}                            & (0.82)        & \dbox{n\texttt{\texttt{/}}a$^{\dag}$}          & (0.28)          & \uwave{0.1$\cdot$$10^{\texttt{-}3}$$^{\dag}$}              & (0.33)         & \dbox{n\texttt{/}a$^{\dag}$}                 &  (0.77)    &  \dbox{n\texttt{/}a$^{\dag}$} & (0.50)  \\[0.09cm] 
    \texttt{bid}        & \underline{\underline{0.1$\cdot$$10^{\texttt{-}1}$$^{\dag}$}}     & (1.46)        & n\texttt{/}a                          & (n\texttt{/}a)  & \uwave{0.3$\cdot$$10^{\texttt{-}3}$$^{\dag}$}              & (0.35)         & 0.1$\cdot$$10^{\texttt{-}4}$          & (0.91)  & 0.1$\cdot$$10^{\texttt{-}4}$ & (0.95) \\[0.09cm]
    \texttt{nreverse}   & 3.1                                                    & (0.30)        & 0.2$\cdot$$10^{\texttt{-}1}$            & (0.22)          & \uwave{0.7$\cdot$$10^{\texttt{-}1}$$^{\dag}$}              & (0.21)         & 20.6                                & (0.27)  & 25.3 & (0.30) \\[0.09cm]
    \texttt{guardians}  & \underline{\underline{15.0}$^{\dag}$}                           & (0.70)        & 0.1$\cdot$$10^{\texttt{-}1}$            & (0.32)          & 0.8$\cdot$$10^{\texttt{-}2}$                      & (0.27)         & 48.6                                & (0.47)  & 59.5   & (0.55) \\[0.09cm]      
    \texttt{crypt}      & \underline{\underline{0.2$\cdot$$10^{\texttt{-}2}$$^{\dag}$}}     & (1.15)        & 0.1$\cdot$$10^{\texttt{-}1}$            & (54.92)         & \uwave{0.4$\cdot$$10^{\texttt{-}2}$$^{\dag}$}              & (0.29)         & 0.1                                 & (0.79)   & 0.2 & (0.72)           
    \\
    \hline
    \texttt{exfilter} & 35.5 & (9.25) & \dbox{n\texttt{\texttt{/}}a$^{*}$} & (3.71)  & n\texttt{/}a & (n\texttt{/}a) & n\texttt{/}a & (n\texttt{/}a)  & n\texttt{/}a  & (n\texttt{/}a) \\[0.09cm]
    \texttt{deepfind} & \underline{\underline{n\texttt{/}a}$^{\dag}$} & (43.48) & n\texttt{\texttt{/}}a & (n\texttt{\texttt{/}}a) & 0.2 & (5.01) & \underline{\underline{n\texttt{/}a$^{*}$}} & (675.89)  & \underline{\underline{n\texttt{/}a$^{*}$}} & (676.56)  \\[0.09cm]
    \texttt{chat-80} & \dbox{n\texttt{/}a$^{*}$} & (334.72) & 0.2$\cdot$$10^{\texttt{-}1}$ & (1,202.55) & \dbox{n\texttt{/}a$^{*}$} & (1,473.90) & \dbox{n\texttt{/}a$^{\dag\dag}$} & (444.72) & \dbox{n\texttt{/}a$^{\dag\dag}$} & (443.65) \\[0.09cm]
    \texttt{LPdoc} & n\texttt{/}a & n\texttt{/}a  & \dbox{n\texttt{/}a$^{\dag}$} & (15.46) & \dbox{n\texttt{/}a$^{*}$} & (19.56) & n\texttt{/}a & n\texttt{/}a & n\texttt{/}a & n\texttt{/}a \\[0.09cm]
    \texttt{Spectector} & \underline{\underline{n\texttt{/}a}$^{\dag}$} & (4.91) & \dbox{n\texttt{/}a$^{\dag}$} & (3.97) & \underline{\underline{n\texttt{/}a$^{\dag\dag}$}} & (4.25) & \underline{\underline{n\texttt{/}a$^{*}$}} & (5.99) & 
\underline{\underline{n\texttt{/}a$^{*}$}} & (6.55)  \\[0.09cm]
    \texttt{s(CASP)} & \dbox{n\texttt{/}a$^{*}$} & (90.18) & \dbox{n\texttt{/}a$^{*}$} & (17.13) & \underline{\underline{n\texttt{/}a$^{**}$}} & (20.43) & \dbox{n\texttt{/}a$^{**}$} & (323.73) & \dbox{n\texttt{/}a$^{**}$} & (271.79)
\botline
  \end{tabular}}
\begin{tablenotes}
\item \errorlegend 
\end{tablenotes}
\end{table}

\newcommand{\lineend}{\\ \hline \ \\ [-5mm] \vspace*{-1mm}}
\begin{table}[!t]
  \centering
  \caption{Details of defects found. The root causes of bugs
    (Class) again include: (I) abstract domain implementation; (II) fixpoint
    algorithms; (III) semantic inconsistencies; (IV) run-time check
    instrumentation; (V) third-party libraries.}\label{bugs}
  \vspace*{-1 mm}
  {\tablefont\begin{tabular}{r@{\hskip 0.08in}l@{\hskip 0.07in}c@{\hskip 0.09in}p{0.8\linewidth}}
    \hline
    \ \\ [-5mm]
    \texttt{\#} & Status & %
                           Class & Description \\ [-2mm]
\hline
1 & New & I & Analysis does not evaluate correctly the \texttt{arg(X,Y,Z)} predicate due to absence of an abstract description in the \textit{def} domain for the case when \texttt{Z} is ground.  \lineend
2 & New & III & The name \textit{covered} is used in two different properties with different semantics.  \lineend %
3 & New & IV & Incorrect run-time semantics for property \texttt{mshare/1} due to being sensitive to variable ordering. 
 \lineend
4 & New & I & Analysis with the \textit{shfr\texttt{+}nonvar} domain of a program containing \texttt{length(X,Y)}, where \texttt{X} is not a variable marks \texttt{length/2} as failing. %
    \lineend
5 & New & I & Abstract definition of the \texttt{\textbackslash=/2} built-in not implemented correctly in the \textit{shfr\texttt{+}nonvar}, \textit{\textit{shfr\texttt{+}clique}}, and \textit{\textit{shfr\texttt{+}clique\texttt{+}def}} domains.  \lineend
6 & New & I & Abstract definition of the \texttt{\textbackslash==/2} built-in not implemented correctly in the \textit{shfr\texttt{+}nonvar}, \textit{\textit{shfr\texttt{+}clique}}, and \textit{\textit{shfr\texttt{+}clique\texttt{+}def}} domains.  \lineend
7 & New & III & The analysis and run-time semantics of \texttt{linear/1} are inconsistent.
 \lineend
   8 & New & IV & The \texttt{mshare/1} implementation for run-time checking considers all variables at a program point. This becomes problematic when both \texttt{clique/1} and \texttt{mshare/1} are present at the same program point. When variables are ``transferred'' from \texttt{mshare/1} to \texttt{clique/1}, \texttt{mshare/1} should
   not consider these variables.%
 \lineend
9 & New & IV & Missing \texttt{clique/1} property implementation for run-time checking.  \lineend
10 & New & I & When analyzing with \textit{\textit{shfr\texttt{+}clique\texttt{+}def}}, the output introduces spurious variables when inferring \texttt{ground/1}.  \lineend
11 & Known & I & Problem in types domains with \texttt{findall/3} calls.  \lineend
12 & Known & I & Problem in types domains with \texttt{setof/3} calls.  \lineend
13 & New & III & Run-time checking instrumentation only understands parametric types that use type symbols as arguments.  \lineend
14 & New & V &
Error in type inference due to skipping testing whether predicates declared as regular types are indeed regular types.  \lineend
15 & New & V & Error in \textit{polyhedra} analysis due to not checking whether variables are numeric. \lineend
16 & New & IV & Run-time implementation of \textit{polyhedra} properties (\texttt{constraint/1}) must check if variables are instantiated: merely evaluating the (in)equality is not sufficient.  \lineend 
17 & New & I & The \textit{greatest lower bound} (\texttt{glb/3}) operation is not defined correctly in \textit{\textit{depth-k}} abstract domain.   \lineend
18 & New & IV & Missing run-time check implementation for \texttt{instance/2}.  \lineend
19 & Known & I & Type analysis failing to generate types for data/dynamic predicates.  \lineend
20 & Known & I & 
\texttt{=/2} built-in not defined correctly in \textit{polyhedra} abstract domain.  \lineend
21 & New & II & Bug introduced in the fixpoint algorithm during code refactoring to incorporate a new transformation aimed at optimizing set-sharing-based analyses. 
\lineend
  \end{tabular}}
\end{table}

\subsection{Results}
\label{sec:results}

\paragraph{\textbf{Cost of the technique.}}
While run-time overhead is not our primary focus, we have evaluated
this aspect in order to study whether the algorithm %
has an acceptable cost. %
This cost obviously has two components: the analysis time and the
testing time (the transformation time is negligible).
Regarding the testing time, note that the execution time of run-time
tests can be reduced significantly through caching
techniques~\citep{cached-rtchecks-iclp2015,rv2014}. However, we decided not
to use these optimizations for a number of reasons: to simplify the
implementation; to avoid dependence on the implementation of other
parts; and to avoid any bugs that optimizations could potentially
hide, making it harder to identify them.
The results are presented in
Tables~\ref{benchmarkstimes1},~\ref{benchmarkstimes2},~\ref{benchmarkstimes3}
and~\ref{benchmarkstimes4} for the different benchmarks and domains.
The experiments were run on a MacBook Air with the Apple M1 chip and
16 GB of RAM.
Each column in these tables corresponds to an abstract domain and in
turn contains two sub-columns.
The first sub-column is the absolute execution time in seconds for
each benchmark once checkification has been applied,
for the properties inferred by the corresponding abstract domain.
This time includes both the transformation process, in which the annotations
of the analysis are modified by replacing the status of \textit{true} assertions
by \textit{check} status and inserting \textit{run-time checks},
as well as the testing time of the instrumented program.
The numbers in parentheses in the second sub-column provide the analysis
times in seconds for the benchmarks, again for the different abstract
domains.

Regarding the testing times, when an error is found during a test run,
the time to do so is typically negligible.  In these cases, we report
instead the testing time after fixing the analysis so that no bugs are
detected and the testing runs to completion. This is obviously also
the case when no bug is detected to begin with.  Thus, the testing
times reported are always for \emph{complete runs}, which we feel are
more useful for estimating the testing cost.

If the time is labeled as \textbf{n/a}, it indicates that there is no
time recorded due to the presence of a timeout or an unresolved bug, or
that no analysis results were available. The latter can be caused by a crash
during analysis or by the analysis output being malformed, e.g., missing
assertions.

The results show that execution time of tests is quite reasonable,
typically taking no more than around 60 seconds.
Performance can be improved if needed by activating only the run-time
semantics of the predicate assertions and/or
disabling multi-variance.

\paragraph{\textbf{Errors found.}}

We now turn our attention to the most important point of whether the
technique can indeed find errors in the analyses.
We have manually analyzed the root causes of the errors found
and classified them into different categories. 
The classification is indicated in
Tables~\ref{benchmarkstimes1} to~\ref{benchmarkstimes4} 
by surrounding the numbers with different patterns. 
These bug categories include (I) defects in the implementation of the
abstract domain, 
(II) defects in the implementation of the fixpoint algorithms, (III)
semantic inconsistencies 
between components of the framework,
(IV) run-time check instrumentation
issues, and (V) defects related to third-party libraries.
Table~\ref{bugs} provides a summary of identified bugs. 
The ``Status'' column indicates whether the bug is new or it was
already a known issue at the time of running the experiments.
The ``Class'' column lists the bug category, while the ``Description'' column
provides a more detailed description of each bug.
We will discuss stylized %
examples from each bug category
in Section~\ref{examples}.
In each column, bugs are additionally marked with symbols
(e.g., $\dag$) to group those that correspond to the same issue.

The results of the experiments conducted so far are promising,
allowing us to draw several significant conclusions and observations.

First and foremost, a good number of bugs and inconsistencies were found
using the technique.
These bugs were quite diverse,
illustrating the power of the algorithm in finding all sorts of issues
of different nature.
\mhnote{Self-note: some useful reflections for us in commented out
  paragraph here.}
No bugs were found for the most mature domains.
On the other hand, bugs were indeed found in the more experimental and 
prototype domains, i.e., domains which were only
partially developed and/or
they or their run-time tests 
supported only a subset of the language at the time of the
experiments (e.g., the \textit{polyhedra} abstract domain).
This included for example no support being available for certain
built-in operations, or no run-time behavior being defined for some
properties used by such domains.
All this has greatly helped complete and strengthen these less mature
domains, since most of the bugs found have now been fixed.
One thing to take into account when reading the results in the tables
is that sometimes many reported
errors are due to just one bug. For example, almost all errors reported
with the \textit{shfr\_clique\_def} domain refer to the same bug found
that affected the analysis of several benchmarks (see 
the markings in the tables mentioned before for other examples). Also, 
it is important to note that 
each program detects a maximum of one bug at a time, since the process halts when
it detects a run-time error.

Another conclusion from the experiments is that benchmark
selection is very important when
testing
specific domains, since each example uses different
\textit{built-ins} and library predicates, exhibits different properties, etc.
Nevertheless, while complex examples such as \texttt{witt} have
resulted most
capable at identifying a wide range of bugs across different domains,
a drawback is that for very complex benchmarks the analysis using the
least mature domains is more prone to fail, and then sometimes no bug is detected.

As mentioned before, in addition to standard benchmarks,
we have also applied our tool to real-life applications. 
In this case, rather than generating test cases, the experiment
consists in compiling the application using all the 
source files as transformed by the algorithm and using the application as
usual, e.g., for the LPdoc documenter, generating with it full
manuals. 

An interesting observation is that the benchmark program suite 
showed similar effectiveness to the real-world  
applications in exposing errors, i.e., most bugs identified in real applications
had already been detected by the set of smaller benchmarks. 
Also, the presence of duplicate bugs is indicative that the algorithm
can identify %
problems consistently.

Finally, the fact that some bugs have already been fixed suggests that
identifying the source of the detected errors in the implementation is relatively
straightforward. 

\subsection{Further Discussion of the Bugs Detected}
\label{examples}

This section illustrates further some of the defects found by the proposed
technique during our experiments reported in Section~\ref{experiments}.
The programs presented are representative of the actual code fragments that
triggered the detection of the bugs; 
however, we generally show simplified and distilled versions for
brevity and
clarity, since the actual code would need significant context to be
understood.
We divide the discussion according to the different classes of bugs (I
to V) from~Section~\ref{experiments}. The bug numbers refer to
Table~\ref{bugs}.

\paragraph{\textbf{Abstract domain implementation} {\normalfont(Class I)}.}
As mentioned before, our testing technique can be seen as
a sanity or coherence check, and thus
it can be targeted to test different
components of the system depending on which ones are assumed to be
trusted.
In general, the \ciao abstract interpretation engine
(the \emph{fixpoint algorithms} and all the surrounding infrastructure
of the system, into which the domains are ``plugged-in'') includes the
components of the analyzer we trust most since they have been used and
refined for
a long time. 
Thus, it makes sense to start by taking this as
the trusted base and aiming the error-finding task at the
different abstract domains.  This %
makes sense specially since \ciaopp is at the same time a production
and a research tool, and new domains are constantly being developed.
\begin{example}[\texttt{Bug \#4}: %
  Missing assertion in library predicate description]
Abstract descriptions of the behavior of the \textit{built-ins}
and %
library predicates are provided for each abstract domain,
either in the file(s) defining the domain, in the libraries themselves, or in both. 
A first use of our algorithm is in order to find errors in these 
specifications and in the implementations of these \textit{built-ins}
and library predicates for a given abstract domain.
In particular,  if during testing a run-time error is found in a program-point
assertion right after a call to a \textit{built-in}, this can be due 
to an error in either the abstract description of that builtin
or in its actual implementation.  
For instance, the code of the \texttt{length/2} library predicate includes
the following (simplified) assertions: %
\prettylstciao
\begin{lstlisting}
:- pred length(+list,-int) + det.
:- pred length(-list,+int) + det.
:- pred length(+list,+int) + semidet.
:- pred length(?,?) + nondet.
\end{lstlisting} 
\noindent
where \texttt{int/1} is a primitive property (in this case a primitive
type) and \texttt{list/1} is defined in a library by the standard
list predicate, and also declared to be a property (in particular, a
regular type) in the usual way:
\prettylstciao
\begin{lstlisting}
:- regtype list/1.
list([]).
list([_|T]) :- list(T).
\end{lstlisting} 
During the tests, in a call to \texttt{length/2}, illustrated by the
following code:
\prettylstciao
\begin{lstlisting}[escapechar=@]
  :- entry p(+list(num),+num,-).
  
  p(X,N,Y) :-
      length([N|X],Y).
\end{lstlisting}
\noindent
i.e., a call with the second argument uninstantiated, the following
analysis output was obtained:
\prettylstciao
\begin{lstlisting}[escapechar=@]
  :- entry p(+list(num),+num,-int).
  
  p(X,N,Y) :-
      true(var(Y), ground([X,N])),
      length([N|X],Y),
      true(unreachable).
\end{lstlisting}
\noindent
i.e., the abstract interpreter (wrongly) inferred failure or error in
the call to \texttt{length/2} and thus that the point after that would
be unreachable. The checkification algorithm translated this output
into:
\prettylstciao
\begin{lstlisting}[escapechar=@]
  :- entry p(+list(num),+num,-int).
  
  p(X,N,Y) :-
      check(var(Y), ground([X,N])),
      length([N|X],Y),
      check(unreachable).
\end{lstlisting}
\noindent
During run-time testing the \texttt{check(unreachable)} literal was
actually reached and executed, which threw the corresponding error.
This led to detecting that the last declaration in
the abstract description for \texttt{length/2} (i.e., ``\texttt{:-
  pred length(?,?) + nondet.}'' in the set of assertions for
\texttt{length/2} above), %
had been deleted by mistake while refactoring some analyzer code.
That assertion is the only one that allows calling \texttt{length/2}
with a variable in the second argument (in fact, in both arguments).
\end{example}

\begin{example}[\texttt{Bug \#10}: Combined abstract domain outputs unknown variables]

While experimenting with the
\textit{CLIQUE-Sharing\texttt{+}Freeness}\texttt{+}\textit{Def}
combined domain an error was detected which can be illustrated with
the following code:

\prettylstciao
\begin{lstlisting}
   :- entry p(+num,-).
 
  p(X,Y) :-
     Y=X.
\end{lstlisting}
\noindent
for which \ciaopp inferred the following information:

\begin{minipage}{\linewidth} %
\prettylstciao
\begin{lstlisting}
   :- entry p(+num,-).
 
  p(X,Y) :-
      true(ground([X,_A])),
      Y=X,
      true(ground([X,Y,_B])).
\end{lstlisting}
\end{minipage}
\noindent
where \texttt{\_A} and \texttt{\_B} were superfluous variables
that have the first occurrence in the \texttt{true(ground(...))}
literals, and thus cannot be ground.
A run-time error was thus produced by the checkified program in the
\texttt{check(ground([X,\_A]))} call, which made us realize that the
\textit{CLIQUE-Sharing\texttt{+}Freeness}\texttt{+}\textit{Def}
abstract domain was not implementing correctly the combination of analyses.
In many combined domains, the combination typically reuses the 
component analyses. For example, all abstract functions for
\textit{CLIQUE-Sharing\texttt{+}Freeness}\texttt{+}\textit{Def} 
initially compute results for the
\textit{sharefree\_clique} and \textit{def} functions. 
Then, they compose the information from
the different domains, eliminating redundancies
over the information inferred by each analysis~\cite{comdom-toplas}.
This is performed by the \textit{reduce} functions, which essentially
compute the \emph{reduced product}. The problem detected stemmed from an
accidental reversal of the sequence of operations in the implementation. 
At first, the abstract function of \textit{def} was executed. However,
the \textit{reduce} function was applied before running
the abstract function of \textit{sharefree\_clique}.
The problem with this is that in some abstract functions a \textit{renaming}
of variables is performed. This \textit{renaming} substitution
replaces each variable in the term it is applied to with distinct fresh variables.
If the \textit{reduce} function is applied before the \textit{renaming} substitution,
it incorrectly treats some variables as distinct when, in fact, they are
identical before renaming.
\end{example}

\paragraph{\textbf{Fixpoint algorithms} {\normalfont(Class II)}.}
Another possible application of the approach is for testing the
abstract interpretation engine (the \emph{fixpoint algorithms} and all
the surrounding infrastructure of the framework)
instead of the domains. This can be done by using domains that are
simple or developed enough to be used as a trusted base.
While the classic fixpoint algorithms are quite stable, new fixpoint algorithms
or modifications of existing fixpoint algorithms 
are sometimes added to the system. Some recent examples include 
a new %
modular and incremental fixpoint~\citep{intermod-incanal-2020-tplp}
and new program transformations
that speed-up set sharing-based analyses by reducing the number of variables
in abstractions~\citep{abstrim-tplp}.
Clearly, these new contributions may introduce new bugs into the system.

A first abstract domain that is useful for this type of checks is the
\emph{concrete domain} itself.
To this end, we
give the analysis a singleton set of initial states as entry point,
i.e., a concrete value, and the analyzer then behaves as a (tabling)
interpreter for the program, starting from the entry point as initial
concrete state.
This test will then detect if the analyzer
incorrectly marks reachable parts of the program as unreachable.

\begin{example}[\texttt{Bug \#21}: Code refactoring]
A code refactoring in the implementation of~\cite{abstrim-tplp} introduced a bug in the handling of
built-ins and cuts. While no issues were identified in the initial version of
the code, the refactored version (where most changes involved renaming operations
and variables) revealed a problem when reapplying the algorithm.\footnote{
  The technique has been included as a fuzz testing component of the \ciao system.
  While the overhead is relatively acceptable for fuzzing tasks, it
  can be less 
  suitable for continuous integration, which requires building tests after each commit.
}
Specifically, the analyzer incorrectly inferred that some predicates containing
built-ins and cuts were dead code, although they were indeed reachable in the concrete
domain.
\end{example}

\paragraph{\textbf{Semantic inconsistencies between components of the framework} {\normalfont(Class III)}.}

Even if every part of the %
system is validated separately,
our tool can still help find inconsistencies among these parts. %
Most components in our system interact via assertions and thus the
semantics of the assertions and of the properties used in the
assertions need to be consistent across all parts.
An interesting case that can occur is when there is a mismatch between
the way properties are understood by the analyzer and the actual
definition of the property that is used in the run-time checks.
Checkification helps detect such inconsistencies.
\begin{example}[\texttt{Bug \#7}: Wrong \texttt{linear/1} run-time semantics]
When the analyzer outputs \texttt{linear(X)}, the semantics is that \texttt{X} is
the list of all program variables that analysis can guarantee to be bound to 
linear terms at the program point, i.e., that the terms that each 
variable in \texttt{X} is bound to do not contain any repeated variables.
The property was inferred correctly by \ciaopp. 
In particular, for a substitution such as \texttt{\{X/f(A,B),
  Y/g(A,C)}\}, \ciaopp was inferring correctly \texttt{true(linear([X,Y]))}.
However, after the checkification conversion to \texttt{check(linear([X,Y]))}, 
a run-time error was being thrown.
This allowed us to notice that in the implementation of the property as a
run-time check
the wrong predicate had been used and what was checked instead of
linearity was that the terms in the list did not share any variable,
which then failed in this case since they share variable \texttt{A}.
\end{example}

\begin{example}[\texttt{Bug \#13}: Different representation of parametric types]
This problem was found when analyzing with type domains.  Assume that the
analyzer has inferred, using, e.g., the \texttt{eterms} domain, that
after success of a call to \texttt{p(X)}, the argument \texttt{X}
is bound to a list of \texttt{a}'s. This is expressed in the output
as:
\noindent
\begin{minipage}{\linewidth} %
\prettylstciao
\begin{lstlisting}[escapechar=@] 
  ...,
  p(X)  
  true(list(rt1, Xs)). 
      
  :- regtype rt1/1.
  rt1(a).
\end{lstlisting}
\end{minipage}
This posed no problems with the run-time checking.  However, since
this style of output can sometimes be verbose, for readability the
user can optionally switch the output so that simple cases are
expressed using quoting (\verb+^/1+) as follows:
\prettylstciao
\begin{lstlisting}[escapechar=@]
  ...,
  p(X)  
  true(list(^(a), Xs)).
\end{lstlisting}
At some point the output had been switched to this format by default
and then we discovered that when executing the
\verb+check(list(^(a), Xs))+ the run-time check instrumentation
did not implement this abridged syntax correctly. This was another 
inconsistency between a possible representation(s) of the abstraction
and what was understood by the run-time checks. 

\end{example}
\paragraph{\textbf{Inconsistencies between properties and their specialized run-time checks} {\normalfont(Class IV)}.}
\mhnote{I am having difficulties distinguishing between Class III and
  Class IV. I added ``specialized'' in title to distinguish them
  more. Please comment. I still have a feeling that some Class IV
  could also be Class III.
  DF: The difference is that in Class III, the error doesn’t come from a specific
  component but from each one interpreting a property’s semantics differently.
  In Class IV, the error is caused by problems with the run-time implementation.}
This is a special case of the previous class, related to the fact that 
it is possible to write an alternative version of a property to be
used specifically in run-time checks, while also keeping the general
definition which may perhaps be easier to read or to be understood by
the static analyzer.

\begin{example}[\texttt{Bug \#9} and \texttt{\#18}: Undefined run-time behavior]
  \label{ex:undefrtbehavior}
In a few instances test case failures stemmed from the specialized run-time behavior
of some properties simply being declared but undefined. For example, testing flagged that 
the implementations for run-time checking of the comparatively less-used \texttt{clique/1} and
\texttt{instance/2} properties were missing.
\end{example}

\begin{example}[\texttt{Bug \#8}: Incompatibility between run-time behaviors]
The approach also detected more subtle issues that only arise when
several properties appear together.
While addressing the issue of the previous example, another
problem was detected when the \texttt{clique/1}
and \texttt{mshare/2} properties appeared in the same assertion.
Set sharing  domains approximate all possible variable sharing
(aliasing) that occurs at a given program point. The property
typically used to denote such sharing among variables is 
\texttt{mshare/1}, where the argument contains a set of sets of
variables.
For instance, let \texttt{V} = \texttt{\{X, Y, Z, W\}} be the set of
variables of interest, normally the variables of the clause.
Consider the abstraction $\lambda =$
\texttt{\{\{X\}, \{X,Y\}, \{X,Y,Z\}, \{X,Z\}, \{Y\}, \{Y,Z\}, \{Z\},
  \{W\}\}}.
Here a set \texttt{\{X,Y\}} represents that, in the terms that
\texttt{X} and \texttt{Y}  are bound to at run time, there may be
variables that appear in both \texttt{X} and \texttt{Y}; 
\texttt{\{W\}} that there may be a variable that appears only in
\texttt{W}; etc. This $\lambda$ will be expressed in the analyzer output as:
\texttt{true(mshare([[X], [X,Y], [X,Y,Z], [X,Z], [Y], [Y,Z], [Z],
  [W]]))}.
We can see that for variables \texttt{\{X, Y, Z\}} 
no information (i.e., top) has been inferred. Indeed for
this set of variables any aliasing may be 
possible since there may be run-time variables shared by any pair of the
three program variables, by the three of them, or not shared at all.
The idea of the \textit{clique} property is to use a more compact
representation for abstractions that contain a powerset, including 
as a widening when abstractions are too large. This has
been shown to pay off in practice~\citep{shcliques-padl06}.
In our example, the \textit{clique}
that will convey the same information with respect to \texttt{\{X, Y, Z\}}
as the sharing set $S = \wp({\{X, Y, Z\}})$ is simply
\texttt{clique([X,Y,Z])}. 
The elements of $S$
are then eliminated from the full sharing set $\lambda$,
since the %
clique makes them redundant.
However, when checkifying the output, it was detected that the
specialized \texttt{mshare/1} run-time checking implementation was still
considering all the variables in the clause (\texttt{\{X, Y, Z, W\}}
in our case). Therefore, \texttt{X}, \texttt{Y}, and \texttt{Z},
were incorrectly being tested for groundness since they are in no
sharing set,  
and thus are interpreted as being ground.
Clearly, this is not what the analysis
inferred.
The \texttt{mshare/1} run-time test was modified to
stop considering %
such variables.
\end{example}

\begin{example}[\texttt{Bug \#3}: \texttt{mshare/1} sensitivity to
  variable ordering]
This case involved a subtle bug in the specialized run-time check for the sharing
abstract domain.  As mentioned before, \texttt{mshare([X,Y])} means
that \texttt{X} may share variables with \texttt{Y} at run time.
Sharing is a symmetric property which does not depend on %
variable ordering, i.e., \texttt{mshare([X,Y])} has the same
meaning as \texttt{mshare([Y,X])}.
Some cases were detected in which a run-time checking 
 error was flagged even though the \texttt{mshare/1} property inferred was 
correct.  The problem was found in an update of the specialized run-time
definition of \texttt{mshare/1}, which introduced an optimization which
made the result sensitive to the ordering of variables.
\end{example}

\begin{example}[\texttt{Bug \#16}: Incorrect definition of \texttt{constraint/1}]
  The \texttt{constraint/1} property is used by some numerical (e.g., polyhedra-related) 
domains. Its argument is a list of linear (in)equalities
that relate variables and integer values.
However, the specialized run-time check of this property was incorrect: it 
checked that the constraints were valid but also whether the arguments were
variables; instead it should have been checking that they were numerical values.
\mhnote{Note to ourselves: this makes me doubt a bit whether constraint perhaps is
  the right name for this property in a (C)LP system.}

\end{example}

\paragraph{\textbf{Integration Testing of the Analyzer with 
    Libraries and Third-Party tools} {\normalfont(Class V)}.}
We have also used the approach to conduct integration tests of the
\textit{Regular Types Library} (an independent \ciao bundle) and check
its correct use within the system.  This library implements
fundamental
operations and procedures for regular types, such as type inclusion,
equivalence, union, intersection, widening, simplification, etc., as
well as storing and manipulating regular types. These operations are
used in \ciaopp domains, analyses, and program transformations.

\begin{example}[\texttt{Bug \#14}: Regular types library type equivalence/simplification bug]
The \texttt{eterms} domain includes the inferred regular types in the
analysis output. In this process, \texttt{eterms} performs type
equivalence and simplification operations in order to 
present the results to users in the most readable form possible. 
The following (simplified) fragment is from one of the test programs:
\prettylstciao
\begin{lstlisting}
  :- prop repeat_elem_list/1 + regtype.
  
  repeat_elem_list([]).
  repeat_elem_list([X,X|Xs]) :-
      num(X),
      repeat_elem_list(Xs).
  
  :- pred diff_elem_list(X) : list(num,X) .
  
  diff_elem_list([]).
  diff_elem_list([X,Y|Xs]) :-
      X =\= Y,
      diff_elem_list(Xs).
\end{lstlisting}
\noindent  
Here \texttt{repeat\_elem\_list/1} is (incorrectly!) declared as a regular
type.
Checkification detected that, in the analysis output,
\textit{eterms} erroneously inferred that \texttt{diff\_elem\_list/1}
produces a \texttt{repeat\_elem\_list/1} upon success:

\prettylstciao
\begin{lstlisting}
  ...
  diff_elem_list([]).
  diff_elem_list([X,Y|Xs]) :-
      check(num(X), num(Y), list(num,Xs)),
      X=\=Y,
      check(num(X), num(Y), list(num,Xs)),
      diff_elem_list(Xs),
      check(num(X), num(Y), repeat_elem_list(Xs)).
\end{lstlisting}
\noindent
This inference is obviously incorrect. 
This allowed us to notice that after some changes
the \textit{Regular Types Library} was failing to check if properties
declared to be regular types were actually regular types, 
which consequently affected the inference of types.
\end{example}

In addition to detecting problems due to the integration with libraries,
the checkification approach also detected incorrectness due to the 
integration with different 
external or third party solvers which can be used by the analyzer.
For example, the \textit{polyhedra} domain uses the \textit{Parma Polyhedra Library} (PPL) as
back-end solver for the handling of numeric approximations.
Using this domain we can detect errors stemming from the \ciao-PPL
integration.

\begin{example}[\texttt{Bug \#15}: Parma Polyhedra Library]
We found that \textit{polyhedra} did not properly handle
some non-numerical parts of programs.
A term like \texttt{X = Y} is translated into the constraint $1\times\texttt{X}-1\times\texttt{Y}=0$ when
analyzing with the abstract domain.
If \texttt{X} and \texttt{Y} are numerical variables, they satisfy the equality dynamically.
However, we discovered that the analysis did not take into account the types of \texttt{X}
and \texttt{Y}, i.e.,
regardless of whether \texttt{X} and \texttt{Y} were numerical or non-numerical,
the analysis treated them as numerical values, not considering their actual types. 
\end{example}

\paragraph{\textbf{Debugging Trust Assertions and Custom Transfer Functions.}}

Finally, the approach has also been useful in finding errors in other
aspects of the framework. As an example, a feature of \ciaopp is that
its analyses can be guided by the user by providing the analyzer with
information that can be assumed to be true at points where otherwise
the analysis would lose precision.  We have already mentioned one of
these mechanisms, \textit{entry} assertions,
which allow providing information on the entry
points to the module being analyzed (i.e., on the calls to the
predicates exported by the module). %
Entry assertions are a special case of \emph{trust} assertions. 
In addition to guiding the analyzer, \emph{trust} assertions are used
to define custom abstract transfer functions, similar to those that need to be
implemented for abstracting each \textit{built-in} within each
domain. \emph{trust} assertions allow the user to do this for any
predicate. Checkification can be used to detect errors in
these assertions. 

\begin{example}[Defining language semantics]
  Let us consider this (again, distilled) part of the specification of
  the exponentiation predicate, where we use the \textit{trust}
  assertion in line~\ref{trust} to state that an integer to the
  power of an integer yields an integer:
\prettylstciao
\begin{lstlisting}[escapechar=@]
  :- trust pred exp_op(+int,+int,-int).@\label{trust}@

  exp_op(X,Y,Z) :- Z is @`@**@'@(X,Y).
  
  exp_list([],_Y,[]).
  exp_list([X|Xs],Y,[Z|Zs]) :- exp_op(X,Y,Z), exp_list(Xs,Y,Zs).
\end{lstlisting}
\noindent
However, in the \textit{ISO-Prolog} standard, the outcome of this
operation is always a floating-point number.  When analyzing this
code, \ciaopp trusts the assertion previously defined and infers that
the result of the operation in the \texttt{exp\_list/3} predicate is
an integer. In the checkified version: 

\prettylstciao
\begin{lstlisting}
  ...
  exp_list([],_Y,[]).
  exp_list([X|Xs],Y,[Z|Zs]) :-
      exp_op(X,Y,Z),
      check(int(X), int(Y), int(Z)),
      exp_list(Xs,Y,Zs).
\end{lstlisting}
\noindent
when we execute a test case with two integers an error is flagged due to this
discrepancy, since the implementation correctly produces a float.
This is a practical application of the proposed algorithm since even a
completely sound analyzer can produce unsound results if it assumes
some assertion to be true when it is not, and thus there will
always be the need to test such properties.
\end{example}
In addition to being able to detect errors in \textit{trust}
assertions, the approach can also be used to find errors in the
mechanisms used internally by the analyzer to apply \textit{trust}
assertions, if we assume instead the information provided in the
\textit{trust} assertion to be correct.

\section{Other Related Work}

\label{sec:related-work}

\selfnote{Self note: \cite{ESEC-FSE-2019-ZhangSYZPS},
  \cite{10.1145/3293882.3330553} and \cite{10.1145/3238147.3240464}
  have all a good related work section for reference}

In this section we mention other related work in addition to the
references interspersed throughout the previous sections.
The fact that the reliability of program analyzers has become crucial
as they have become increasingly practical and widely adopted in
recent years, is now widely recognized~\citep{symex-transf:fse-ni-13}.
This has led to significant research interest recently. %

On the formal verification side,
there have been some pen-and-paper proofs, such as that of the Astrée
analyzer~\citep{98e19ed52e984f7e8c48c14a16a2bfa2},
some automatic and interactive proofs, such
as~\cite{10.1007/3-540-44659-1_9, 10.1145/565816.503293}.
As mentioned in the introduction, a recent relevant effort has been
aimed at verifying the partial correctness of the \ciaopp analysis
algorithm (also referred to as the ``top-down solver''), using the
Isabelle prover~\citep{Top_Down_Solver-AFP,Paulson90}, but mechanical
verification of the actual implementations remains a challenge.
An approximation to this problem is \cite{10.1007/978-3-642-38856-9_18}
and \cite{10.1145/2676726.2676966} who 
have developed and proved the soundness
of a static analyzer, extracting several abstract domains and a
fixed-point verifier directly from Coq formalizations.

In the context of testing static analyzers
significant work has
been done. \cite{10.1145/3238147.3240464} and
\cite{dblp:journals/stvr/midtgaardm17} use mathematical properties
of abstract domains as test oracles. Specifically,
\cite{10.1145/3238147.3240464} employ such properties
to validate the soundness and precision of numerical
abstract domains, while \cite{dblp:journals/stvr/midtgaardm17}
apply that approach to checking type analysis
using QuickCheck~\citep{Claessen:2000:QLT:351240.351266-icfp}.

The closest works to ours are those
that cross-check dynamically observed and statically inferred
properties~\citep{10.1145/2491411.2491439, ESEC-FSE-2019-ZhangSYZPS,
  10.1007/978-3-642-28891-3_12, 10.1145/3088515.3088521}.
In~\cite{10.1145/2491411.2491439} the actual pointer aliasing
in concrete executions is cross-checked with the pointer aliasing inferred by an
aliasing analyzer.
They are also able to deal
with multi-variance and path-sensitivity.
Compared to us, they require significant
tailored instrumentation which cannot be reused for testing other
analyses. However, their approach is agnostic to the (C) aliasing
analyzer.
Another cross-check is done in~\cite{ESEC-FSE-2019-ZhangSYZPS} for C
model checkers and the \textit{reachability} property, but they obtain
the assertions dynamically, and check them statically, complementarily
to our approach. %
Unlike us, they again need tailored instrumentation that
cannot be reused to test other analyses, and their benchmarks must be
deterministic and with no input, the latter limiting the power of the
approach as a testing tool. However, their approach is agnostic to the
(C) model checker.
In~\cite{10.1007/978-3-642-28891-3_12} a wide range of static analysis
tests are performed over randomly generated programs as input. Among others,
they check dynamically, at the end of the program, one assertion
inferred statically, and they perform the sanity check to ensure
that the analyzer behaves as an interpreter when run from a singleton
set of initial states.
Our approach differs from these previous works
by identifying issues throughout the entire static analyzer framework,
rather than focusing on specific components.
In this setting, \cite{10.1145/3293882.3330553} propose an
automatic technique to evaluate soundness of program analyzers based
on differential testing. From seed programs, they generate program analysis
benchmarks and compare the overall frameworks of software model checkers.
Differential analysis presents a significant challenge, requiring
multiple analyzers with identical input/output behavior to
be applied. A potential problem arises if one or more analyzers
behave differently, making it difficult to determine which
is correct.
\cite{he2024finding} apply two oracles to compare static analyzers.
The first oracle is constructed using dynamic program executions.
A random program generated by Csmith~\cite{10.1145/2737924.2737986},
a generation-based C fuzzer,
serves as test input,
and its dynamic execution result (e.g., a run-time error on some program path)
is used as a test oracle to validate the static analyzer.
The second oracle is a static oracle based on metamorphic
relations.
The approach involves selecting a conditional expression 
of some statement in the program as a target and generating
an equivalent boolean expression.
The static analyzer should produce identical truth values for these
equivalent expressions, any discrepancy indicates a potential defect in the
analyzer.

There is work on testing program analyzers in
other domains. For instance, \cite{10.1145/1670412.1670413}
introduce a grammar-based fuzzer to identify crashes in SMT
solvers, using delta debugging to minimize generated instances. 
Additionally, \cite{symex-engine-tester:ase17} combine random program generation
with differential testing to evaluate symbolic execution engines.
~\cite{10.1145/1287624.1287651} implement Java programs as test oracles
for abstract syntax trees for testing refactoring engines.
There is also considerable work in testing compilers.
\cite{10.1145/2594291.2594334,10.1145/2983990.2984038,10.1145/2814270.2814319}
apply Equivalence Modulo Inputs (EMI) testing, which involves
mutating unexecuted statements of an existing program under certain
inputs to produce new equivalent test programs. 
\cite{10.1145/2254064.2254104} propose a test-case reduction
technique to construct minimal inputs that trigger compiler bugs.
Finally,
Building on Csmith~\cite{10.1145/1993498.1993532},
\cite{10.1145/2737924.2737986} introduce CLsmith, which designs
six modes to generate test programs.

We argue that, in comparison with these related proposals, our
approach provides a solution that is simpler, elegant, more general,
and easier to implement, specially when the different system
components are integrated as in the Ciao assertion model.

\section{Conclusions}

\label{conclusions}

We have proposed and studied \emph{checkification}, an
automatic method for testing %
static analysis tools by checking
that the properties inferred statically are satisfied
dynamically.
A fundamental strength of our approach lies in its simplicity, which
stems from framing it within the \ciao assertion model and using its
assertion language.
We have shown how checkification can be implemented 
with comparatively little effort by combining the static
analyzer, the run-time checking framework, the random test case
generator, and the unit-test framework, 
together with a reduced amount of glue code.
This code implements the proposed
algorithm, and pilots the combination and interplay of the intervening
components to effectively implement the overall approach.

Following this approach, we have constructed a quite complete
implementation of the checkification method 
and applied it to testing a large number of the abstract
interpretation-based analyses in \ciaopp, representing different
levels of code maturity, as well as to the framework itself, the
interaction with libraries, the interaction with third-party code,
etc. In the study we analyzed both standard benchmarks and
real-world tools.
We applied our technique not only to 
state properties such as variable sharing/aliasing, modes, linearity,
numerical properties, types, and term structure, but also to
computational properties such as determinacy and (non)failure.
As we discussed, the technique should also be applicable to
other, more complex, computational properties
such as cost and even termination,
although they obviously bring in specific challenges and theoretical
limits for run-time checking.

The experimental results %
show that our tool can effectively
discover and locate
interesting, %
non-trivial and previously undetected bugs, with reasonable overhead,
not only in the less-developed parts of the system but also in corner
cases of the more mature components, such as the handling of
\textit{built-ins}, run-time checking instrumentation, etc.
The approach has also proven useful for detecting issues in auxiliary
stages of analysis and verification, including assertion simplification, pretty
printing, abstract program optimizations and transformations, etc.
We have also observed that it is generally not too hard to locate the
source of the errors from the information produced by the tests, and
as a result the vast majority of the detected issues were in fact
fixed during the experimental study.
While testing approaches are obviously ultimately insufficient for
proving the correctness of analyzers, and thus it is clearly
worthwhile to also pursue the avenue of code verification, we believe
that our results show that the checkification approach can be a
practical and effective technique for detecting errors in large and
complex analysis tools.

\bibliographystyle{tlplike}
\bibliography{extracted}

\end{document}